\documentclass[a4paper,11pt]{article}
\usepackage{jheppub}
\usepackage[utf8x]{inputenc}
\usepackage{graphicx}
\usepackage{float}
\usepackage[T1]{fontenc}
\usepackage{epsfig}
\usepackage{color}
\usepackage{tikz}
\usepackage{amsmath,mathtools}
\usepackage{amssymb}
\usepackage{pgfplots}
\usepackage{subfloat}
\usepackage{amsfonts}
\usepackage{braket}
\usepackage{cleveref}
\usepackage{epstopdf}
\usepackage{caption}
\usepackage{subcaption}
\usepackage{enumitem}
\usepackage{etoolbox}
\usepackage{orcidlink}
\usepackage{comment}
\graphicspath{ {./Figures/} }
\usepackage[titletoc,toc,title]{appendix}
\usetikzlibrary{fillbetween}
\usepackage{hyperref}
\hypersetup{
	colorlinks=true,
	linkcolor=blue,
	filecolor=red,      
	urlcolor=cyan,
	citecolor=red
}

\title{Perturbative study of Supercritical Crossover in  Noncommutative-corrected Spacetime}

\author[a]{Ankit Anand\orcidlink{0000-0002-8832-3212}}
\author[b,*]{\;and\;Shoucheng Wang\orcidlink{0000-0002-5909-6233}}

\affiliation[a]{
	Department of Physics, Indian Institute of Technology Kanpur, 208016, India}
    \affiliation[b]{
	School of Science, Hunan Institute of Technology, Hengyang 421002, China.}

\emailAdd{anand@iitk.ac.in}
\emailAdd{scwang@hnit.edu.cn}

\let\oldthefootnote\thefootnote
\renewcommand{\thefootnote}{}
\footnotetext{*Corresponding author.}
\let\thefootnote\oldthefootnote

\abstract{We analytically study the Widom line and supercritical crossover of noncommutative charged AdS black holes. Treating the noncommutative parameter $\alpha$ perturbatively, we compute thermodynamic quantities and the scaled variance $\Omega$ in both canonical and extended ensembles. The Widom line is identified as the extremum of $\Omega$. Using a Landau expansion near the critical point, we derive the two symmetric crossover branches $L^{\pm}$, which obey $\delta T\sim \left|\Delta Q\right|^{\beta+\gamma}$, $\delta S\sim \left|\Delta Q \right|^\beta$ in the canonical ensemble and $\delta P\sim \left|\Delta T\right|^{\beta+\gamma}$, $\delta \rho\sim \left|\Delta T\right|^{\beta}$ in the extended ensemble. These scaling relations conform to the mean‑field universality class ($\beta=1/2$, $\gamma=1$), and the noncommutative parameter only shifts subleading amplitudes without altering the universality class. Numerical verification and complete supercritical phase diagrams are also presented using supercritical crossover lines. Our results show that noncommutative corrections preserve the mean‑field universality of black hole supercriticality.
}

\begin{document}

\maketitle

\section{Introduction}

The study of supercritical phenomena has become an important theme across many areas of physics. Classically, the supercritical regime—lying beyond the critical point of a first-order phase transition—was regarded as a homogeneous state in which the distinction between liquid and gas vanishes \cite{10.1063/PT.3.1796}. Recent investigations, however, demonstrate that this picture is incomplete: gas-like and liquid-like features persist in the supercritical domain and can be identified via thermodynamic response functions, such as Widom lines \cite{Xu_2005, Ruppeiner_2012, PhysRevLett.112.135701, PhysRevE.95.052120, Gallo2014, Li_2024, 10.1063/1.1671624, RJFLeotedeCarvalho_1994, PhysRevE.51.3146, Bolmatov2013, doi:10.1021/acs.jpcb.9b04058, Simeoni2010, Simeski2023}, or through dynamical probes like Frenkel lines \cite{Yoon_2018, PhysRevLett.111.145901, Prescher_2017, Bolmatov_2015, Fomin_2018, Fomin2015, PhysRevE.85.031203, 2023PhRvR...5a3149H, jiang2024experimentalobservationgappedshear}. Strikingly, such crossover behavior has been observed not only in ordinary fluids (e.g., water) but also in microscopic systems such as quantum chromodynamics (QCD) \cite{Sordi:2023cjq, Li_2024, PhysRevLett.75.1040, Stephanov:2008qz, Stephanov:2011pb, Jim_nez_2021}. This naturally motivates the question of whether analogous supercritical phenomena may emerge in gravitational systems, particularly in the thermodynamics of black holes.

Black holes, since the pioneering work of Bekenstein and Hawking \cite{Hawking:1982dh}, are known to obey a consistent set of thermodynamic laws, in which the horizon area corresponds to entropy and the surface gravity to temperature. In Anti-de Sitter (AdS) backgrounds, the interpretation of the cosmological constant as a thermodynamic pressure \cite{Kubiznak:2012wp, Kubiznak:2016qmn, Kastor:2009wy} reveals a rich phase structure, including the Hawking-Page transition and van der Waals-type criticality~\cite{Cvetic_1999, Cvetic:1999rb, Kubiznak:2012wp}. This correspondence strongly suggests that black holes can access supercritical regimes analogous to those in conventional matter, where crossover behavior between distinct thermodynamic states may occur. Despite this, most existing work has focused on critical phenomena near the transition point, with comparatively little attention paid to developing a unified description of the supercritical sector \cite{DasBairagya:2019nyv, Sahay:2017hlq, Wei:2019yvs, Wei:2023mxw}.

At the same time, the possibility that spacetime itself acquires a noncommutative structure at short distances has emerged as a compelling feature of candidate quantum gravity theories. Noncommutative geometry, originally motivated by string theory where the antisymmetric $B$-field induces noncommuting coordinates on D-branes \cite{15}, provides a natural framework to encode such effects. In this setting, classical point-like sources are replaced by smooth Gaussian or Lorentzian distributions \cite{160}, which can regularize curvature singularities and alter the effective mass and charge profiles of black holes. These modifications introduce controlled quantum corrections into black hole thermodynamics, offering a tractable avenue to investigate the interplay between microscopic geometry and macroscopic gravitational phenomena. Embedding noncommutative effects into the AdS black hole phase structure, therefore, provides a promising framework to study how quantum geometry reshapes criticality, stability, and the possible emergence of supercritical crossover behavior. Despite substantial progress in understanding near-critical black hole thermodynamics~\cite{wei2020critical, altamirano2013reentrant, gunasekaran2012extended}, a systematic study of the supercritical region, particularly in the presence of quantum-geometric modifications, remains largely unexplored. Establishing such a framework is both timely and essential, with implications ranging from holography and strongly coupled field theories~\cite{maldacena1998largeN, hartnoll2008holographic, witten1998ads, aharony2000largeN, gubser1998gauge, witten1998confinement} to gravitational-wave phenomenology and cosmology~\cite{abbott2016gw150914, abbott2021gwtc2, domenech2021cosmogw, gonin2025pbh}.

The noncommutative parameter $\alpha$ is treated perturbatively, under the assumption $\alpha \ll 1$. This is justified because noncommutative effects are expected to arise predominantly at Planckian scales, contributing as small quantum corrections to the classical black hole geometry \cite{nicola2010nc, smeared2006nc}. Perturbative treatment allows a systematic evaluation of leading-order modifications to thermodynamic quantities, phase structure, and supercritical crossover lines, without the need to solve the full noncommutative field equations, which are typically analytically intractable \cite{smailagic2003fuzzy}. Importantly, this approach ensures that the classical limit is smoothly recovered as $\alpha \rightarrow 0$, while providing a controlled framework to study how quantum spacetime corrections influence black hole critical behavior \cite{nozari2008nc}.

The study of critical phenomena in many-body systems often relies on the mean-field approximation, in which collective interactions are effectively replaced by their averaged effects. This approach, originally developed in statistical mechanics \cite{stanley1971phase, goldenfeld1992lectures}, provides a powerful framework to capture the essential features of continuous phase transitions. One of its central outcomes is the concept of universality, which asserts that diverse physical systems exhibit identical critical exponents and scaling laws near criticality, regardless of microscopic details.  In black hole thermodynamics, a wide range of analyses have demonstrated that charged and rotating AdS black holes share the same critical exponents as the van der Waals fluid, namely $\bar{\alpha} = 0, \beta = \tfrac{1}{2}, \gamma = 1, \delta = 3$ in exact agreement with mean-field predictions \cite{kubiznak2012pvn, gunasekaran2012extended, wei2019critical}. Here, the critical exponent is denoted by $\bar{\alpha}$ so that the reader should not be confused with the noncommutative parameter $\alpha$; henceforth, $\alpha$ denotes only the noncommutative parameter. This universality highlights the robustness of black hole phase transitions and motivates the use of analytic mean-field methods to probe beyond the critical point, including the supercritical crossover regime and the Widom line.

In this work, we analytically analyze the scaling behavior of a noncommutative black hole in the supercritical regime using the Landau expansion. Meanwhile, a more concrete approach to supercritical crossover has been developed to capture the supercritical scaling behavior, based on the identification of two distinct crossover lines $L^{\pm}$ that emerge from the critical point and extend into the supercritical region \cite{Wang:2025ctk}. Unlike the Widom line, which provides a single symmetric curve, the \(L^{\pm}\) lines capture the asymmetric thermodynamic response on either side of the critical isochore and are defined by the loci of maximal response functions—such as the isothermal compressibility in the extended phase space or the specific heat in the canonical ensemble—along paths parallel to the critical isochore. These lines obey universal scaling laws in the context of mean field theory,
\begin{equation}
\delta P^{\pm} \propto (T-T_{\rm c})^{\beta + \gamma}\,,
\label{eq:H_scaling}
\end{equation}
and 
\begin{equation}
\delta \rho^{\pm} \propto (T-T_{\rm c})^{\beta}\,,
\label{eq:eta_scaling}
\end{equation}
where $T$ is the control field, $\delta P^{\pm}$ and $\delta \rho^{\pm}$ denote deviations of ordering field and order parameter from the critical isochore, and $\beta, \gamma$ are the mean-field critical exponents. This structure has been verified in a variety of black hole systems, including the $\text{RN–AdS}_4$ black hole in both extended and non-extended ensembles, hairy black holes, higher-dimensional and Gauss–Bonnet black holes, as well as Barrow black holes with fractal corrections (see the updated version of \cite{Wang:2025ctk}). The $L^{\pm}$ lines thus provide a systematic and universal framework for characterizing supercritical thermodynamics. In this work, we extend this analysis to noncommutative black holes, computing the $L^{\pm}$ lines considering the perturbative effects of the noncommutative parameter $\alpha$, and demonstrating that the universal scaling structure persists. The influence of the noncommutative parameter on the supercritical phase diagram has also been visualized in a concrete manner. These two approaches complement each other in demonstrating the supercritical scaling behavior of charged black holes in the presence of noncommutative corrections.

\section{Review of Noncommutative Black Holes}
\label{sec:Review of Noncommutative Black Holes}

Noncommutative geometry provides a natural framework for incorporating minimal length effects expected from quantum gravity. In the context of black hole physics, two broad and inequivalent approaches have been developed to implement spacetime noncommutativity. The first is an algebraic approach, where noncommutativity is introduced through deformations of the spacetime coordinate algebra, typically realized via Drinfel'd twists and the Seiberg--Witten map. The second is an effective geometric approach, in which point-like matter sources are replaced by smooth, spatially smeared distributions.

The algebraic deformations of spacetime symmetries, based on Drinfel'd twists and the Seiberg--Witten map \cite{Seiberg:1999vs}. In this framework, noncommutativity is encoded directly at the level of spacetime coordinates, leading to modified field equations and geometric structures. Recently, this approach has been employed to construct noncommutative dyonic black hole solutions in Einstein gravity coupled to nonlinear electrodynamics \cite{Bokulic:2025mtd}. In the twist-induced formulation, one begins with a classical static and spherically symmetric black hole background and introduces noncommutative corrections perturbatively in a deformation parameter $\alpha$, assumed to be small. To leading order, the corrected spacetime metric have off diagonal term depending on the model-independent function encoding the leading noncommutative correction \cite{Bokulic:2025mtd}. The emergence of an off-diagonal component reflects the intrinsic anisotropy introduced by the twist deformation. Importantly, conserved quantities such as the mass and electromagnetic charges remain unchanged at this order, while the spacetime geometry itself acquires genuine noncommutative corrections.

The another description of noncommutative black holes is based on the smearing of point-like mass and charge distributions over a minimal length scale, leading to regularized geometries and perturbative corrections to thermodynamic quantities. This construction has proven particularly suitable for exploring black hole thermodynamics, critical phenomena, and supercritical crossover behavior, as demonstrated in the subsequent analysis of Widom lines and crossover scaling. Nevertheless, it is important to emphasize that this effective realization represents only one possible implementation of spacetime noncommutativity. Recently, it has been shown that in \cite{Hadri:2025mvu, Anand:2025ab}, by choosing the Lorentzian smeared mass and charge densities, the metric function is
\begin{eqnarray} 
f(r) &=& 1+\frac{32 \alpha  M}{\pi ^2 \alpha ^2+64 r^2}-\frac{4 M}{\pi  r} \tan ^{-1}\left(\frac{8 r}{\pi  \alpha }\right)+\frac{4 Q^2}{r^2} \left(\frac{\tan ^{-1}\left(\frac{8 r}{\pi  \alpha }\right)}{\pi }-\frac{8 \alpha  r}{\pi ^2 \alpha ^2+64 r^2}\right)^2 + \frac{r^2}{\ell^2} \ , \nonumber \\
&=& \underbrace{1-\frac{2 M}{r}+\frac{Q^2}{r^2}+\frac{r^2}{\ell^2} }_{\stackrel{}{\mbox{$f_{RN}(r)$}}} \;+\;\alpha \left( \frac{M}{r^2}-\frac{Q^2}{r^3} \right) \;+\; \alpha^2 \left( \frac{Q^2}{4 r^4} \right) \;+\; \mathcal{O}(\alpha^3) \ .
\end{eqnarray}
Since we are studying charged black holes within a noncommutative framework, noncommutativity naturally appears as a deformation of the classical Reissner–Nordström (RN) black hole. It is obvious to expect that the thermodynamic quantities are also corrected by a non-commutative parameter.

\subsubsection*{Thermodynamics}

This identification helps in writing the non-commutative corrected quantity as
\begin{eqnarray}
    \mathcal{W}_{\rm NC} = \mathcal{W}_{\rm RN}+\alpha \Delta \mathcal{W} + \mathcal{O}(\alpha^2)  \ ,
\end{eqnarray}
where $\mathcal{W}$ can be mass, temperature, or critical quantities of the equation of state, etc. Additionally, this deformation helps preserve the RN structure while introducing perturbative corrections, which smoothly interpolate between the standard point-source limit at large distances and a regularized core at short scales. In the physically relevant regime \(\alpha \ll r_h\), with \(r_h\) denoting the horizon radius, the smearing effects are exponentially suppressed, and the geometry asymptotically reduces to the usual RN solution. It is therefore natural to interpret noncommutativity as a perturbative deformation of the RN spacetime.

From a thermodynamic perspective, this deformation implies that the thermodynamic quantities, equation of state, and associated critical quantities can be computed systematically as an expansion in powers of \(\alpha\). The leading-order terms reproduce the standard RN–AdS results, while higher-order corrections encode deviations induced by spacetime noncommutativity. The first-order correction to the thermodynamic quantities is
\begin{eqnarray}\label{Hawking_Temp}
    \Delta M = \frac{1}{4}  \left(\frac{r_h^2}{l^2}-\frac{Q^2}{r_h^2}+1\right) \quad;\quad  \Delta T = \frac{-1}{8 \pi } \left(\frac{1}{l^2}-\frac{3 Q^2}{r_h^4}+\frac{1}{r_h^2}\right) \quad ;\quad \Delta S_{BH} = \pi r_h \ , \nonumber
\end{eqnarray}  
where the metric retains a purely diagonal form and noncommutativity manifests through effective modifications of the mass and charge profiles. We show explicitly these corrections play a crucial role in shaping the Widom line and the analytic crossover behavior in the supercritical regime. The Widom line analysis presented in this work demonstrates that noncommutative effects preserve the underlying mean-field scaling while inducing systematic shifts in the crossover slopes and amplitudes.


\subsection{Review of the mean-field paradigm and response functions}

Critical phenomena exhibit a remarkable degree of universality across diverse physical systems. Mean-field theory provides a powerful and economical framework for capturing this universality, relying on symmetry principles and analyticity rather than on microscopic details. Its applicability to black hole thermodynamics is particularly compelling, since the underlying degrees of freedom of quantum gravity remain largely unknown. The mean-field description serves as an effective macroscopic theory for black hole phase transitions, including those arising in non-commutative gravity.

Near a critical point, the thermodynamic behavior may be characterized by an effective order parameter $\phi$, which measures deviations from equilibrium.  The associated thermodynamic potential admits a Landau-type expansion,
\begin{equation}
\mathcal{F}(\phi,T) = \mathcal{F}_0(T) + \frac{1}{2}a(T)\phi^2  + \frac{1}{3}b\phi^3 + \frac{1}{4}c\phi^4 - h\phi + \cdots ,
\end{equation}
where the coefficient $a(T)$ changes sign at the critical temperature $T_c$, while higher-order terms ensure thermodynamic stability. The quantity $h$ denotes the field conjugate to $\phi$, encoding the response to external perturbations. Equilibrium configurations follow from minimizing the thermodynamic potential, leading to the mean-field equation of state
\begin{equation}
a(T)\phi + b\phi^2 + c\phi^3 - h = 0 \ .
\end{equation}
This nonlinear relation governs the order parameter's response to its conjugate field and captures the essential features of critical behaviour.

A central diagnostic of criticality is the susceptibility, $\chi$, which measures the sensitivity of the order parameter to infinitesimal variations of the conjugate field. As the system approaches the critical point, the susceptibility diverges according to the mean-field scaling law
\begin{equation}
\chi \sim |\tau|^{-\gamma} \quad\text{where} \quad
\tau = \frac{T-T_c}{T_c} \ .
\end{equation}
This divergence reflects the flattening of the thermodynamic potential near its minimum and the consequent amplification of fluctuations.

A defining feature of mean-field theory is the universality of its critical exponents. 
For a wide class of systems, including black holes, these exponents take the classical values
\begin{equation*}
\bar{\alpha} = 0 \qquad;\qquad  \beta = \frac{1}{2} \qquad ;\qquad \gamma = 1 \qquad;\qquad  \delta = 3 \ .
\end{equation*}
In non-commutative black hole models, the presence of a fundamental length scale modifies the thermodynamic coefficients and shifts the critical parameters.  However, the scaling relations believed to be governed by these exponents typically remain unchanged, highlighting the robustness of the mean-field description.

Although phase coexistence terminates at the critical point, the influence of criticality persists into the supercritical regime. This remnant structure is organized by the Widom line, which generalizes the notion of critical behavior beyond equilibrium phase transitions. Rather than being associated with divergences, the Widom line is defined as the locus of maximal thermodynamic response. The useful quantity for identifying this is the scaled variance $\Omega$, which measures the strength of thermodynamic fluctuations in the one-phase region. Formally derived from second derivatives of the Gibbs free energy, $\Omega$ may be viewed as a generalized susceptibility that remains finite away from the critical point while developing pronounced maxima near the Widom line. In black hole thermodynamics, the identification of the relevant susceptibility depends on the chosen ensemble. In non-extended ensembles with fixed charge or angular momentum, response functions such as the specific heat provide analogous measures of thermodynamic sensitivity, while in the extended phase space, the thermodynamic volume serves as the order parameter, and the isothermal compressibility acts as the corresponding susceptibility.

\section{Analytical derivation of Widom Line}\label{Sec:Analytical derivation of Widom Line}

Beyond the critical point, the thermodynamic phase structure of black holes is no longer governed by phase coexistence, yet it retains clear signatures of criticality. The thermodynamic behaviour is less governed by phase coexistence and more by fluctuations. From an information-theoretic standpoint, criticality reflects an enhancement of correlations and a corresponding loss of distinguishability between competing macrostates. While the first-order transition terminates at the critical point, the imprint of this enhanced information flow persists into the supercritical regime as sharply peaked, yet finite, response measures.

A useful framework for capturing this residual critical behaviour is provided by the Widom line, which identifies the locus in the $(Q, T)$ plane for the non-extended ensemble and in the $(P, T)$ plane for the extended ensemble where the response functions attain their maximal values~\cite{PhysRevE.85.031203,Ouyang:2024ckt}. This curve extends into the supercritical regime, distinguishing regions characterized by qualitatively different microscopic structures, analogous to liquid-like and gas-like behaviour in conventional fluids. It is worth studying the effect of the non-commutative parameter $\alpha$ on this. 

We use the idea~\cite{Li_2024, Zhao:2025ecg, Anand:2025rzh} and use the normalized variance \(\Omega\), defined as the ratio of second- to first-order temperature derivatives of the Gibbs free energy. The Widom line goes beyond the coexistence curve into the supercritical region, where the first-order phase transition disappears, and the thermodynamic variables no longer exhibit discontinuities. But they have very rapidly and smoothly transitioned between separate black hole phases, and the persistence of crucial correlations for $T > T_c$. By using a scaled variance $\Omega$, constructed from higher-order derivatives of the Gibbs free energy, the Widom line is then determined by
\[
\left(\frac{\partial \Omega}{\partial T}\right)_P = 0 \ ,
\]
which identifies the temperature of maximal thermodynamic response at fixed pressure. The resulting curve $T_W(P)$ separates liquid-like and gas-like black hole microstructures and provides a smooth extension of critical behavior into the supercritical region. Recently, an alternative definition based on the projection of complex Lee–Yang zeros onto the real phase plane has been proposed in~\cite{Xu:2025jrk}. For the black hole, one starts by writing the Gibbs free energy in terms of the horizon radius $r_+$. The normalized variance can be written as
\begin{eqnarray}
\Omega = \frac{k_2}{k_1} = \frac{G''}{G' T'} - \frac{T''}{(T')^2} \quad \text{where} \quad k_n \equiv \left(\frac{\partial^n G}{\partial T^n}\right)_P \ ,
\end{eqnarray}
which measures the strength of thermodynamic fluctuations. The maxima of $\Omega$ trace the Widom line, marking the supercritical crossover where response functions exhibit pronounced but continuous behavior. This quantity serves as an information-theoretic susceptibility, measuring the relative amplification of fluctuations under infinitesimal thermal perturbations. By construction, \(\Omega\) quantifies the strength of thermodynamic fluctuations and encodes the rapidity with which the system explores nearby macrostates.

In the next subsection, we determine the Widom line perturbatively in the noncommutative parameter, analyzing its behaviour in both the non-extended and extended thermodynamic ensembles.

\subsection{Widom Line in Non-Extended Ensemble}

We analyze the thermodynamic critical behaviour of noncommutative black holes in the canonical ensemble, where the electric charge $Q$ is allowed to fluctuate while the AdS length scale $l$ (equivalently, the pressure) is held fixed. The charge serves as a control parameter governing the phase structure, closely analogous to a field in standard condensed-matter systems. The appropriate thermodynamic potential is the Helmholtz free energy, $F$, whose differential follows directly from the first law of black hole thermodynamics as 
\begin{equation}
 dF=-S\,dT+\Phi\,dQ  \ .
\end{equation}
The Hawking temperature expressed as a function of entropy $S$ and charge $Q$ takes the form
\begin{eqnarray}
T(S,Q) &=& \frac{\pi  l^2 \left(S-\pi  Q^2\right)+3 S^2}{4 \pi ^{3/2} l^2 S^{3/2}}-\frac{\alpha }{2 \pi  l^2}-\frac{\alpha ^2 \left(2 \pi  l^2 \left(S-2 \pi  Q^2\right)+S^2\right)}{16 \left(\sqrt{\pi } l^2 S^{5/2}\right)} \ .
\label{Temp_in_S_Q}
\end{eqnarray}
This characterises the thermal response of the black hole in the canonical ensemble, including perturbative corrections induced by spacetime noncommutativity. We first compute the critical quantities for later use.

At the critical point, the temperature exhibits an inflection in entropy at fixed charge. This is encoded in the conditions
\begin{equation}
\left( \frac{\partial T}{\partial S} \right)_Q = 0  \qquad;\qquad \left( \frac{\partial^2 T}{\partial S^2} \right)_Q = 0 \ .
\label{crit_conditions}
\end{equation}
We now evaluate these conditions perturbatively in the noncommutative parameter $\alpha$. By using Eq.~\eqref{Temp_in_S_Q} and Eq.~\eqref{crit_conditions}, we have to solve 
\begin{eqnarray}\label{T_S_Expansion}
    \frac{3 \pi ^2 l^2 Q^2-\pi  l^2 S+3 S^2}{8 \pi ^{3/2} l^2 S^{5/2}}+\frac{\alpha ^2 \left(-20 \pi ^2 l^2 Q^2+6 \pi  l^2 S+S^2\right)}{32 \sqrt{\pi } l^2 S^{7/2}} &=& 0 \nonumber \\
    -\frac{3 \left(5 \pi ^2 l^2 Q^2-\pi  l^2 S+S^2\right)}{16 \left(\pi ^{3/2} l^2 S^{7/2}\right)}+\frac{\alpha ^2 \left(140 \pi ^2 l^2 Q^2-30 \pi  l^2 S-3 S^2\right)}{64 \sqrt{\pi } l^2 S^{9/2}} &=& 0 \ .
\end{eqnarray}
Our aim is to compute the canonical critical entropy $S_c$ and critical charge $Q_c$ up to second order~\footnote{We are not getting the first order correction because the $\mathcal O(\alpha)$ correction to the temperature is a constant, it does not affect the shape of $T(S)$ and hence cannot shift the inflection point. Only the $\mathcal O(\alpha^2)$ term introduces a nontrivial $S$-- dependence, lifting the degeneracy and producing a shift in $S_c$ and $ Q_c$.} in the noncommutative parameter $\alpha$. For this, we write the critical quantities as 
\begin{eqnarray}\label{T_Q_Expansion}
    T_c = T_0+\alpha \, T_1 +\alpha^2 \, T_2 \qquad ; \qquad Q_c = Q_0+\alpha \, Q_1 +\alpha^2\, Q_2 \ .
\end{eqnarray}
By putting Eq.~\eqref{T_Q_Expansion}, Eq.~\eqref{T_S_Expansion} and solving them simultaneously, we have 
\begin{eqnarray}\label{Critical_in_S_Q}
    S_c  = \frac{\pi  l^2}{6}-\frac{1}{8} (7 \pi ) \alpha ^2 \quad ; \quad Q_c = \frac{l}{6}-\frac{17 \alpha ^2}{24 l} \quad ; \quad T_c = \frac{\sqrt{6}}{3\pi  l}-\frac{\alpha }{2 \pi  l^2}-\frac{5 \alpha ^2}{4 \sqrt{6} \pi  l^3} \ . 
\end{eqnarray}
As expected, the classical RN-AdS critical behaviour is recovered at leading order; spacetime noncommutativity induces systematic perturbative shifts in the critical quantities. These corrections play a crucial role in shaping the Widom line structure.

Now, we express the normalized fluctuation \(\Omega\) as a function of the entropy and charge, expanding systematically in the noncommutative parameter,
{\footnotesize \begin{eqnarray}\label{Omega(S,Q)_expr}
    \Omega(S,Q) &=& \frac{8 l^2\pi ^{3/2}  S^{3/2}}{\pi  l^2\, \left(3 \pi  Q^2-S\right)+3 S^2}-\frac{\alpha ^2 \left[\left(\pi ^3 l^4 Q^2\right) \left(9 \pi  Q^2+S \right)+\left( 3\pi l^2  S^2\right) \left(S-12 \pi  Q^2\right)+3 S^4\right]}{ \left(6 l^2 \pi ^{5/2} \sqrt{S}\right)^{-1} \;\left[\left(\pi  l^2\right) \left(3 \pi  Q^2-S\right)+3 S^2\right]^3}  
\end{eqnarray}}
The Widom line at fixed charge is defined by the extremality condition i.e., $\partial_S \Omega(S,Q)|_{Q}=0$ which we again solve perturbatively by expanding the Widom entropy as
\begin{eqnarray}\label{Widom_entropy_ansatz}
    S_W=S_W^{(0)} + \alpha\, S_W^{(1)} + \alpha^2 \, S_W^{(2)} +\mathcal{O}(\alpha^3) \ .
\end{eqnarray}
By differentiating Eq.~\eqref{Omega(S,Q)_expr}, expanding by putting Eq.~\eqref{Widom_entropy_ansatz} and solving order by order, we have 
\begin{eqnarray}\label{Widom_Entropy_expr}
    S_W = \frac{ \pi  l}{6} \left(\sqrt{l^2+108 Q^2}-l\right)-\frac{\pi  \alpha ^2}{4}  \left(\frac{2 l}{\sqrt{l^2+108 Q^2}}+5\right) \ .
\end{eqnarray}
By putting Eq.~\eqref{Widom_Entropy_expr} into Eq.~\eqref{Temp_in_S_Q} we have 
\begin{eqnarray}
    T_W = \frac{\frac{48 \sqrt{6} Q^2}{\left(l \left(\sqrt{l^2+108 Q^2}-l\right)\right)^{3/2}}-\frac{2 \alpha }{l^2}-\frac{\alpha ^2 \left(-2 l^3+2 l^2 \sqrt{l^2+108 Q^2}+261 Q^2 \sqrt{l^2+108 Q^2}-432 l Q^2\right)}{\sqrt{\frac{l^2}{6}+18 Q^2} \left(l \left(\sqrt{l^2+108 Q^2}-l\right)\right)^{5/2}}}{4 \pi } \ .
\end{eqnarray}

To extract universal near-critical behavior, we expand about the critical charge using \(\Delta Q = Q/Q_c - 1\). By using Eq.~\eqref{Critical_in_S_Q} and retaining terms up to quadratic order and working consistently at \(\mathcal{O}(\alpha)\), the Widom temperature takes the form
\begin{eqnarray}
    T_W = \frac{8 \sqrt{6} l^2-12 \alpha  l-5 \sqrt{6} \alpha ^2}{24 \pi  l^3}-\frac{\Delta Q \left(137 \alpha ^2+16 l^2\right)}{32 \left(\sqrt{6} \pi  l^3\right)}+\frac{\Delta Q^2 \left(1693 \alpha ^2+224 l^2\right)}{256 \sqrt{6} \pi  l^3} \ .
\end{eqnarray}
Finally, introducing the reduced temperature \(\Delta T=(T_W-T_c)/T_c\), the Widom line is finally expressed as
\begin{eqnarray}\label{final_Widom_line}
    \Delta T = \left(-\frac{1}{4}-\frac{\sqrt{6} \alpha  }{16 l} -\frac{153 \alpha ^2 }{64 l^2}\right)\Delta Q + \left(\frac{7}{16}+\frac{1917 \alpha ^2 }{512 l^2}+\frac{7 \sqrt{6} \alpha  }{64 l}\right)(\Delta Q)^2 + \cdots \ ,
\end{eqnarray}
where we have used Eq.~\eqref{Critical_in_S_Q}. Notably, while the noncommutative parameter \(\alpha\) shifts the critical data and the absolute location of the Widom line, it does not alter the reduced functional form at leading order. The supercritical crossover exhibits a symmetric response, encoded entirely by a single Widom curve because of no external ordering field.



\subsection{Widom Line in Extended Ensemble}

Following \cite{Xiao:2023lap, Kastor:2009wy, Dolan:2010ha, Kubiznak:2012wp} and identifying pressure with cosmological constant and by introducing the specific volume ($v$), the critical point associated with the small/large black hole transition is obtained from the inflection point conditions of the \(P\!-\!v\) isotherm. As shown in~\cite{Anand:2025ab}, the analytical computation of critical quantities is complicated, as the resulting polynomial in \(v\) is generically of sixth degree. Consequently, a closed-form solution for \((P_c, T_c, v_c)\) is not feasible. Instead, one may determine the critical quantities perturbatively in \(\alpha\). Interestingly, it was found in~\cite{Anand:2025ab} that no \(\mathcal{O}(\alpha)\) correction appears in \(P_c\); the first modification arises only at order \(\alpha^2\).  Combining all the critical quantities yields the correction in universal ratio at the second order in $\alpha$. So, the noncommutative effects leave an observable imprint on the universal thermodynamic behaviour of charge-noncommutative-corrected black holes.  

For later purpose as discussed in~\cite{Anand:2025rzh}, we introduce $\rho = v^{-1}$, with this the equation of state including the leading $\alpha$ correction
\begin{align}
P(\rho,T)&= -\frac{\rho^2}{2\pi}+ \frac{2 Q^2 \rho^4}{\pi}+ \rho T+ \frac{\alpha}{3\pi}\left(\rho^3 -16Q^2\rho^5 + \pi T\rho^2\right) \ .
\label{eq:eos_simplified}
\end{align}
The critical point $(\rho_c,T_c)$ is defined by the inflection point of the $P$--$\rho$ isotherm,
\begin{equation}
\left(\frac{\partial P}{\partial\rho}\right)_T = 0
\qquad;\qquad
\left(\frac{\partial^2 P}{\partial\rho^2}\right)_T = 0 .
\label{eq:crit_def}
\end{equation}
Taking the derivative of Eq.~\eqref{eq:eos_simplified} with respect to $\rho$ at fixed $T$, we obtain
\begin{align}
0 &=-\frac{\rho}{\pi}+ \frac{8Q^2\rho^3}{\pi}+ T + \frac{\alpha}{3\pi}\left(3\rho^2 -80Q^2\rho^4 + 2\pi T\rho \right) \ , \nonumber \\
0 &= -\frac{1}{\pi} + \frac{24Q^2\rho^2}{\pi} + \frac{\alpha}{3\pi} \left(6\rho -320Q^2\rho^3 + 2\pi T \right) \ .
\label{eq:d2Pdrho2}
\end{align}
Since again we are solving perturbatively in $\alpha$, we determine the critical point perturbatively by expanding
\begin{equation}
\rho_c = \rho_0 + \alpha\rho_1 + \cdots \qquad;\qquad T_c = T_0 + \alpha T_1 +  \cdots \ .
\label{eq:ansatz}
\end{equation}
By putting Eq.~\eqref{eq:ansatz} into Eq.~\eqref{eq:d2Pdrho2} and solving order by order simultaneously, we have
\begin{align}\label{Criticality_in_rho_P}
\rho_c &=  \frac{1}{2 \sqrt{6} Q}+\frac{\alpha }{24 Q^2}  \qquad ;\qquad T_c = \frac{1}{3 \sqrt{6} \pi  Q} - \frac{\alpha }{72 \pi  Q^2} \qquad ;\qquad P_c = \frac{1}{96 \pi  Q^2}-\frac{17 \alpha ^2}{6912 \pi  Q^4} \ .
\end{align}
The leading-order behaviour remains governed by the classical RN–AdS universality class, with noncommutativity manifesting itself as a hierarchy of small corrections to the critical data. Such perturbative modifications accumulate in the supercritical regime and directly influence the location and curvature of the Widom line.

Working in the extended ensemble, we express $\Omega (P,r)$ and write its dependence on noncommutativity through a systematic perturbative series as 
\begin{equation}\label{Omega_Extended}
\Omega(r,P,Q;\alpha) =\underbrace{ \frac{8 \pi  r^3}{8 \pi  P r^4+3 Q^2-r^2}}_{\stackrel{}{\mbox{$\Omega_0$}}}+\alpha \underbrace{\frac{4 \pi   r^2 \left(-8 \pi  P r^4+9 Q^2-r^2\right)}{\left(-8 \pi  P r^4-3 Q^2+r^2\right)^2}}_{\stackrel{}{\mbox{$\Omega_1$}}}+\mathcal{O}\left(\alpha ^2\right) \ .
\end{equation}

The Widom line, the locus of extrema of the $\Omega$. We solve $\partial _r \Omega =0$ to obtain the Widom radius. Again, assuming a perturbative expansion of the Widom radius,
\begin{equation}\label{Ansatz_rw}
r_W = r_0 + \alpha r_1 +\cdots  \ .
\end{equation}
Differentiating Eq.~\eqref{Omega_Extended} w.r.t. $r$ and using Eq.~\eqref{Ansatz_rw} we have 
\begin{eqnarray}
    -\frac{8\pi  r_0^2 \left(8 \pi  P r_0^4-9 Q^2+r_0^2\right)}{\left(-8 \pi  P r_0^4-3 Q^2+r_0^2\right)^2}+\frac{ \alpha\left(3 Q^2 r_0^2 \left(1-96 \pi  P r_0^2\right)+8 \pi  P r_0^6 \left(8 \pi  P r_0^2+3\right)+27 Q^4\right)}{(8 \pi  r_0 (2 r_1+1)^{-1})\left(8 \pi  P r_0^4+3 Q^2-r_0^2\right)^3} = 0 \ . \nonumber
\end{eqnarray}
We solve this at every order in $\alpha$. The zeroth order reduces to the classical RN, and by choosing the positive root, a real horizon radius is ensured. Finally, solving for the first order correction to the Widom radius, we have up to linear correction in $\alpha$ we have 
\begin{equation}\label{Widom_radius}
r_w = \sqrt{\frac{-1 + \sqrt{1 + 288\pi P Q^2}}{16\pi P}} - \frac{\alpha}{2} \ .
\end{equation}
Now, by putting the  widom radius~\eqref{Widom_radius} into Hawking temperature Eq.~\eqref{Hawking_Temp}, by introducing the reduced pressure deviation $\Delta P = \tfrac{P-P_c}{P_c}$, the Widom temperature reduces to 
\begin{eqnarray}\label{Widom_Tempr_in_DeltaP}
T_W &=& -\frac{\alpha -4 \sqrt{6} Q}{72 \left(\pi  Q^2\right)}+\frac{\Delta P \left(3 \sqrt{6} Q-2 \alpha \right)}{144 \pi  Q^2}-\frac{\Delta P^2}{64 \left(\sqrt{6} \pi  Q\right)}+O\left(\Delta P^3\right) + \cdots \ .
\end{eqnarray}
Inverting the critical temperature in Eq.~\eqref{Criticality_in_rho_P}, we have $ Q = \tfrac{\sqrt{1-3 \pi  \alpha  T_c}+1}{6 \sqrt{6} \pi  T_c}$ and putting this in Eq.~\eqref{Widom_Tempr_in_DeltaP} we have 
\begin{eqnarray}\label{DeltaT_Compact}
    T_W = T_c+\Delta P \left(\frac{3 T_c}{8}-\frac{15}{32} \pi  \alpha  T_c^2\right) - \Delta P^2 \left(\frac{9}{256} \pi  \alpha  T_c^2+\frac{3 T_c}{64}\right)+O\left(\Delta P^3\right) + \cdots  \nonumber \\
   \Delta T = \frac{T_W-T_c}{T_c} =  \left(\frac{3}{8}-\frac{15 \pi T_c}{32}  \alpha  \right) \Delta P - \left(\frac{9}{256} \pi  \alpha  T_c+\frac{3}{64}\right) \Delta P^2 + \cdots \ .
\end{eqnarray}
Now, we invert this relation perturbatively; we assume a series expansion of $\Delta P$ in powers of $\Delta T$ as
\begin{equation}\label{DeltaP_ansatz}
\Delta P = c_1\,\Delta T + c_2\,(\Delta T)^2 + \mathcal{O}(\Delta T^3) \ .
\end{equation}
Substituting this ansatz into Eq.~\eqref{DeltaT_Compact}, we obtain
\begin{align}
\Delta T = \left(\frac{3}{8}-\frac{15 \pi T_c}{32}  \alpha  \right) c_1\,\Delta T + \left(\left(\frac{3}{8}-\frac{15 \pi T_c}{32}  \alpha  \right) c_2 - \left(\frac{9}{256} \pi  \alpha  T_c+\frac{3}{64}\right) c_1^2\right)\Delta T^2 + \mathcal{O}(\Delta T^3) \ .
\end{align}
By matching powers of $\Delta T$ order by order, we yield the coefficient in Eq.~\eqref{DeltaP_ansatz} as
\begin{eqnarray}
    c_1= \frac{8}{3}+\frac{10 \pi \,  T_c}{3} \alpha + \cdots \qquad ;\qquad c_2 = \frac{8}{9}+4 \pi  \alpha  T_c + \cdots
\end{eqnarray}
Finally, putting this in Eq.~\eqref{DeltaP_ansatz} we have 
\begin{eqnarray}\label{eq:universal_widom_line}
    \Delta P = \left(\frac{8}{3}+\frac{10 \pi  \alpha  T_c}{3}\right)\Delta T + \left( \frac{8}{9}+4 \pi  \alpha  T_c \right) (\Delta T)^2 + \cdots \ .
\end{eqnarray}
This demonstrates that the Widom line preserves the linear mean-field scaling
\begin{equation}
\Delta P \propto \Delta T \ ,
\end{equation}
while $\alpha$ produces only a perturbative shift in the slope. As expected, the supercritical crossover is symmetric and described by a single Widom line. In the next section, we introduce a conjugate thermodynamic variable that lifts this degeneracy, leading to a bifurcation of the crossover into two separate branches $L^\pm$, which capture finer details of near-critical behavior.

\section{Supercritical Crossover Analysis}

The conjugate control parameter is added; the symmetric supercritical response characterized by a single Widom trajectory is lifted. The resulting symmetry breaking splits the crossover into two distinct supercritical lines, denoted by $L_\pm$, which encode the asymmetric thermodynamic response on either side of the critical isochore. These lines provide a finer resolution of near-critical behaviour. Near the critical point, the black hole system no longer exhibits a true first-order phase transition. Instead, it enters a crossover regime in which thermodynamic response functions remain finite but display sharp extrema. The locations of these extrema define the supercritical crossover lines $L_\pm$, which may be interpreted as smooth continuations of the coexistence curve into the single-phase region. Although the transition is no longer nonanalytic, critical correlations continue to dominate the thermodynamic behaviour and govern the structure of the crossover.

Operationally, we characterize the crossover by tracking the maxima of appropriate susceptibilities along paths parallel to the critical isochore. The choice of susceptibility depends on the ensemble under consideration: in the canonical ensemble it is the specific heat at fixed charge, $C_Q$ while in the extended phase space, the relevant response function is the isothermal compressibility $\kappa_T$. We employ a Landau-type expansion of the thermodynamic potential, in which the order parameter couples linearly to its conjugate field and the leading nonlinear contribution. This structure shows that of ordinary fluids and magnetic systems and implies that the crossover lines obey universal scaling relations governed by mean-field critical exponents. As we demonstrate below, this universality persists even in the presence of spacetime noncommutativity, with noncommutative effects entering only through subleading corrections to the critical amplitudes.



\subsection{Non-extended Ensemble Supercritical Crossover}

We proceed to analyze the supercritical crossover behavior within the non-extended ensemble, where the charge is treated as a fixed thermodynamic variable and fluctuations occur in the entropy–charge sector. In the non-extended ensemble, the entropy \(S\) serves as an effective order parameter, while the electric charge \(Q\) plays the role of a conjugate control variable, closely resembling the field-order-parameter structure encountered in classical mean-field theories. Importantly, although spacetime noncommutativity modifies the underlying geometry, the near-critical thermodynamic response remains governed by universal scaling relations.

To characterize deviations from criticality, we introduce the entropy fluctuation ($\mathbb{S}= S - S_c$) and charge fluctuations($\mathbb{Q} = Q - Q_c$) around the critical point. Near \(T_c\), the temperature variation admits a Landau-type expansion in the order parameter \(\mathbb{S}\),
\begin{equation}\label{deltaT}
    \delta T =  T(S,Q) - T(S_c,Q)= J\,\mathbb{Q}\, \mathbb{S} + K\, \mathbb{Q}^3 + \cdots \ . 
\end{equation}
The coefficient \(J\) governs the linear coupling between charge and entropy fluctuations, while the cubic coefficient \(K\) encodes the leading nonlinear self-interaction responsible for the onset of phase bifurcation. Evaluated at the critical point, these coefficients are given by
\begin{eqnarray}
J = \left.\frac{\partial^2 T}{\partial S \partial Q}\right|_c = \frac{9 \sqrt{6}}{2\pi ^2 l^4}-\frac{81 \sqrt{6} }{16\pi ^2  l^6} \alpha ^2 + \cdots \quad;\quad
K = \frac{1}{6}\left.\frac{\partial^3 T}{\partial S^3}\right|_c = \frac{9 \sqrt{6}}{2\pi ^4 l^7}+\frac{225 \sqrt{6}}{16\pi ^4 l^9}\,\alpha ^2 +\cdots \ .
\end{eqnarray}
While the presence of the noncommutative parameter modifies the numerical values of these coefficients, it does not alter the polynomial structure of the expansion. As a result, the non-extended ensemble continues to exhibit mean-field-type critical behaviour.

The response of the system to charge fluctuations is measured by the susceptibility
\begin{eqnarray}
    \chi_Q = \left( \frac{\partial \mathbb{S}}{\partial \delta T} \right)_Q = \frac{1}{J\mathbb{Q} + 3K \mathbb{S}^2} \ .
\end{eqnarray}
The extrema of \(\chi_Q\) identified the loci of enhanced thermodynamic sensitivity, analogous to compressibility maxima in ordinary fluids. Solving the extremality condition yields two solutions,
\begin{eqnarray}\label{Crossover_Non_extended}
  \mathbb{S}_\pm^2 &=& \left(\frac{\pi ^2 l^3}{3}-\frac{17 \pi ^2 l}{12} \alpha ^2  \right)\,\mathbb{Q}  \quad;\quad
\delta T_\pm  = \pm \left( \frac{6 \sqrt{2}}{\pi  l^{5/2}}-\frac{39}{\sqrt{2} \pi  l^{9/2}}\alpha ^2 \right) \, \mathbb{Q}^{3/2}  \ .  
\end{eqnarray}
These branches, denoted \(L_+\) and \(L_-\), define a pair of supercritical crossover curves that coalesce at the critical point. Together, they form a Widom-like structure in the non-extended ensemble, separating regimes that are continuously connected to liquid-like and gas-like black hole phases. The characteristic scaling \(\delta T_\pm \propto \mathbb{Q}^{3/2}\) is a direct consequence of mean-field universality and persists despite noncommutative deformations of the spacetime background.

\begin{figure}[th]
    \centering
    \includegraphics[scale=0.55]{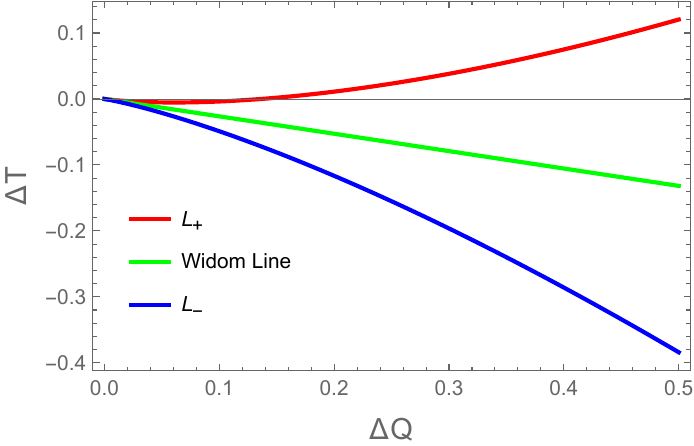}
    \includegraphics[scale=0.55]{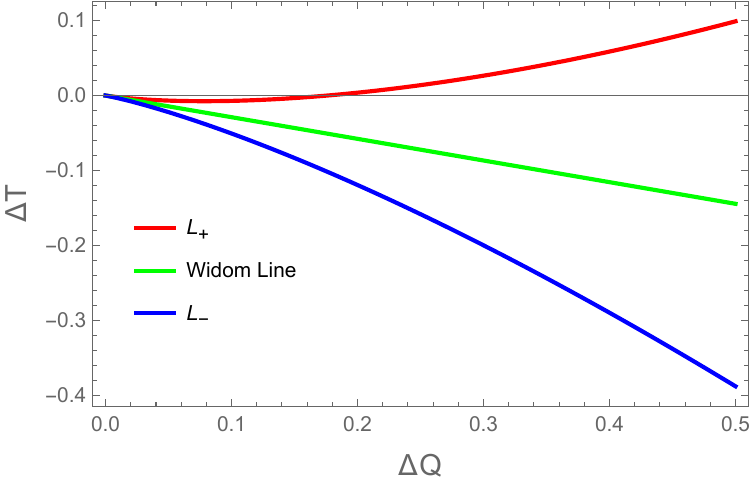}
    \caption{Supercritical crossover structure (\textbf{left : }$\alpha=0.05$ and \textbf{right : }$\alpha=0.1$) in the non-extended ensemble. This illustrates the crossover fan emerging from the critical point, with its central line corresponding to the Widom line.}
    \label{fig:Non-Extended} 
\end{figure}

At this stage, it is important to distinguish between two different notions of temperature variation. The Landau expansion in Eq.~\eqref{deltaT} is expressed in terms of $\delta T \equiv T(S,Q) - T(S_c,Q)$, which differs from the thermodynamic deviation $T - T_c = T(S,Q) - T(S_c,Q_c)$. Since, in the present perturbative setup, $T_c \neq T(S_c,Q)$, one cannot identify $\delta T$ with $T - T_c$. Therefore, to extract the universal scaling associated with the crossover lines, we define different reduced temperature as
\begin{equation}
\widetilde\Delta T \equiv \frac{\delta T}{T_c} \ .
\end{equation}
Substituting Eq.~\eqref{Crossover_Non_extended} and expanding near the critical point $Q \approx Q_c$, we obtain
\begin{equation}
\widetilde\Delta T_\pm = \frac{\delta T_\pm}{T_c}= \pm \left(\frac{1}{\sqrt{2}} +\frac{\sqrt{3}}{4l}\alpha- \frac{69}{8\sqrt{2}l^2}\alpha^2 \right) |\Delta Q|^{3/2} \ .
  \end{equation}
To obtain the full crossover structure, we must additionally include the linear contribution arising from the shift of the critical isochore, $T(S_c,Q) - T(S_c,Q_c)$, which generates the Widom line discussed in Sec.~\ref{Sec:Analytical derivation of Widom Line}. Finally, combining the linear Widom trajectory in Eq.~\eqref{final_Widom_line} with the nonlinear scaling above, the complete crossover lines take the form
  \begin{equation}
  \Delta T_{L\pm}(\Delta Q) =\left(-\frac{1}{4}-\frac{\sqrt{6}}{16l}\alpha-\frac{153}{64l^2}\alpha^2\right)\Delta Q \pm\left(\frac{1}{\sqrt{2}}+\frac{\sqrt{3}}{4l}\alpha-\frac{69}{8\sqrt{2}l^2}\alpha^2\right)|\Delta Q|^{3/2}+ \cdots .
  \end{equation}
Here, the linear term represents the mean-field drift of the thermodynamic trajectory near criticality, while the nonlinear contribution reflects symmetric fluctuations that split the response into two distinct crossover branches. The noncommutative parameter \(\alpha\) shifts the location of the supercritical crossover; it does not alter the reduced functional form at leading order. The supercritical crossover exhibits a fan-like structure, encoded entirely by the $L_\pm$ lines. The corresponding plot of $\Delta T_{L\pm}(\Delta Q)$ in Fig.~\ref{fig:Non-Extended} illustrates the supercritical crossover structure in the non‑extended ensemble for two representative values of the noncommutative parameter, $\alpha=0.05$ (left) and $\alpha=0.1$ (right). The central green line represents the Widom line, given by the linear mean‑field relation $\Delta T =\left(-\frac{1}{4}-\frac{\sqrt{6}}{16l}\alpha-\frac{153}{64l^2}\alpha^2\right)\Delta Q$ at leading order, which smoothly extends the coexistence line into the supercritical region. The red and blue branches, labeled $L_+$ and $L_-$, correspond to the two crossover lines obtained from the extremal condition of the susceptibility; they satisfy the universal scaling $\delta T \sim |\Delta Q|^{3/2}$ near the critical point. The shaded region between the two branches defines the crossover fan, within which thermodynamic response functions exhibit pronounced maxima. Comparing the left and right panels, one observes that increasing the noncommutative parameter leaves the slope of the crossover lines unchanged, meaning that the underlying mean-field universal structure remains unaffected.

\begin{figure*}[t!]
    \centering
    \includegraphics[width=0.45\columnwidth]{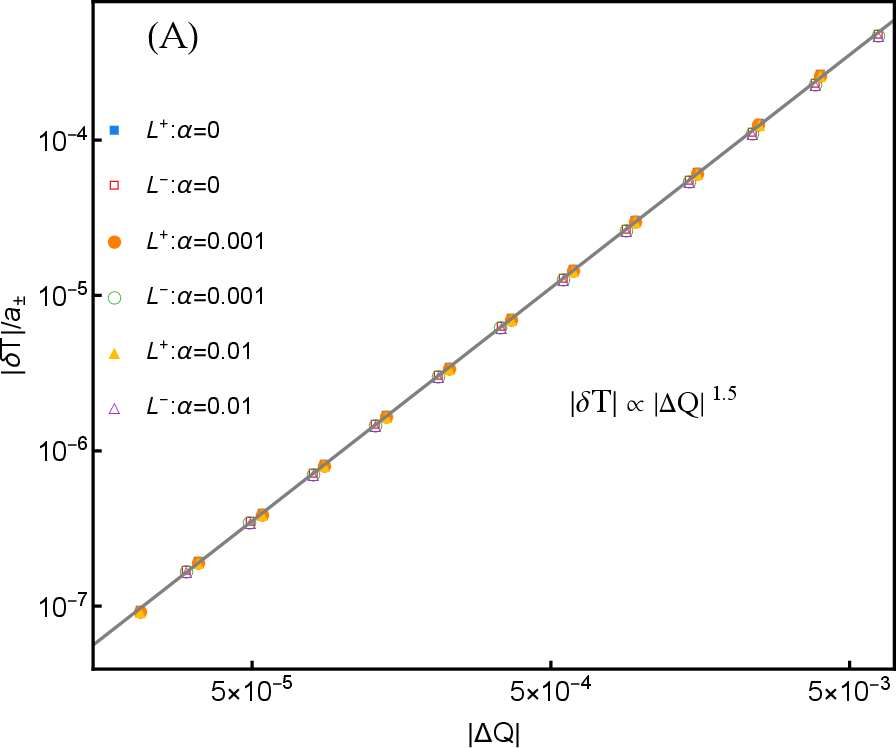}
    \includegraphics[width=0.46\columnwidth]{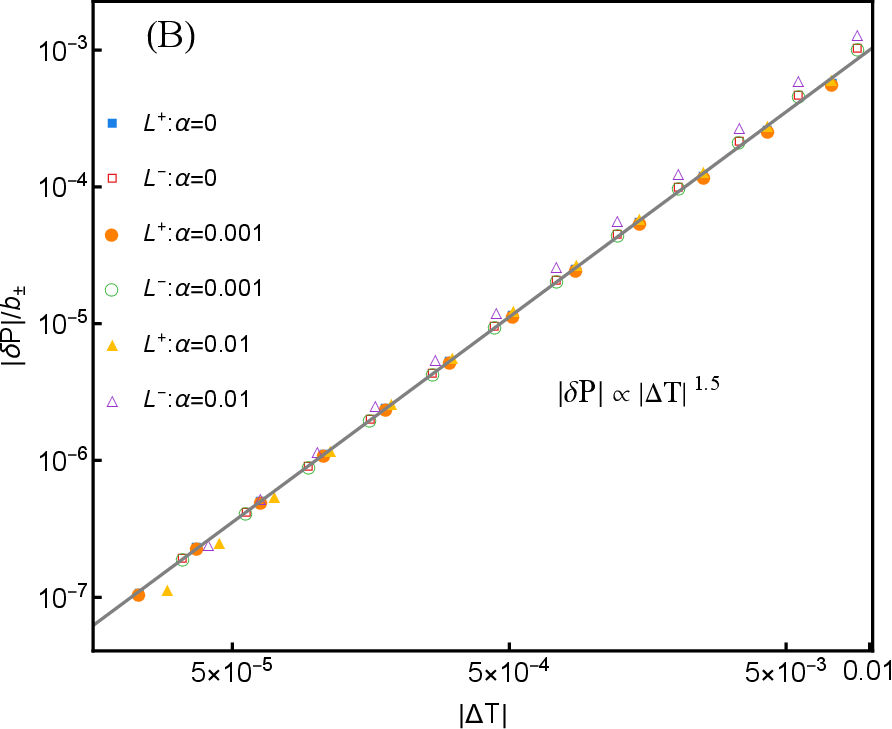}
    \hspace*{-0.3cm}
    \includegraphics[width=0.45\columnwidth]{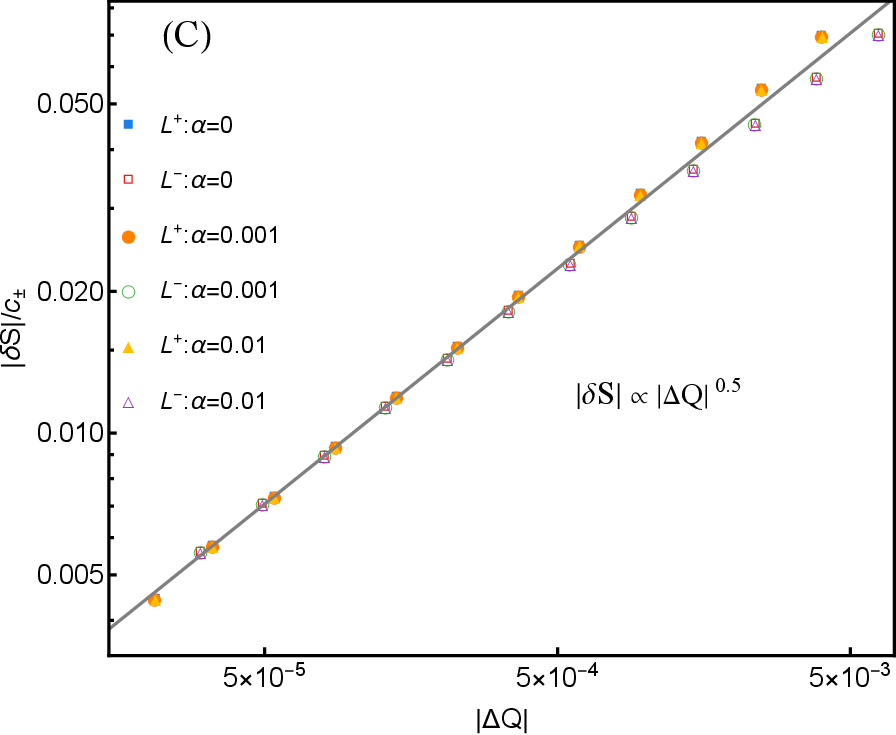}
    \includegraphics[width=0.46\columnwidth]{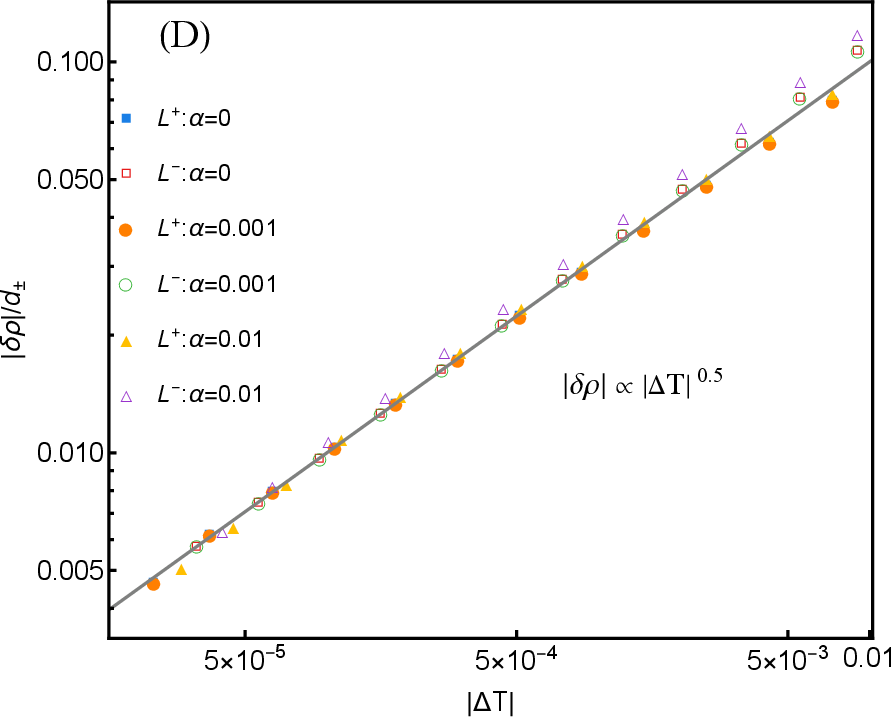}
    \caption{Scaling law near the critical point of (A)(B) the external field, Eq.~\eqref{eq:H_scaling}, and (C)(D) the order parameter, Eq.~\eqref{eq:eta_scaling}, along the supercritical crossover lines, $L^+$ and $L^-$ (open and filled symbols correspondingly), for non-extended (A)(C) and extended (B)(D) phase space respectively. The noncommutative parameter is set to three representative values: $\alpha=0, 0.001, 0.01$. The values of $a_\pm$ to $d_\pm$ are summarized in Table~\ref{tab:coefficients}.}
    \label{fig:scaling}
\end{figure*}

\begin{figure}[th]
    \centering
    \includegraphics[scale=0.42]{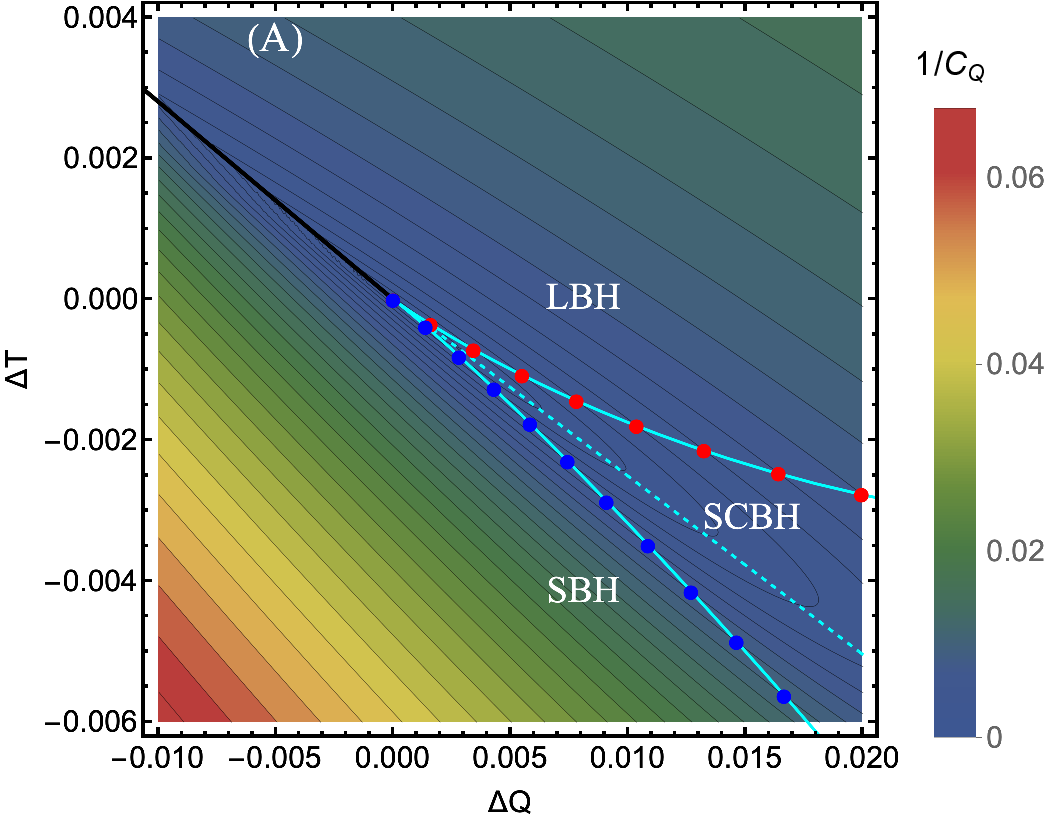}
    \includegraphics[scale=0.42]{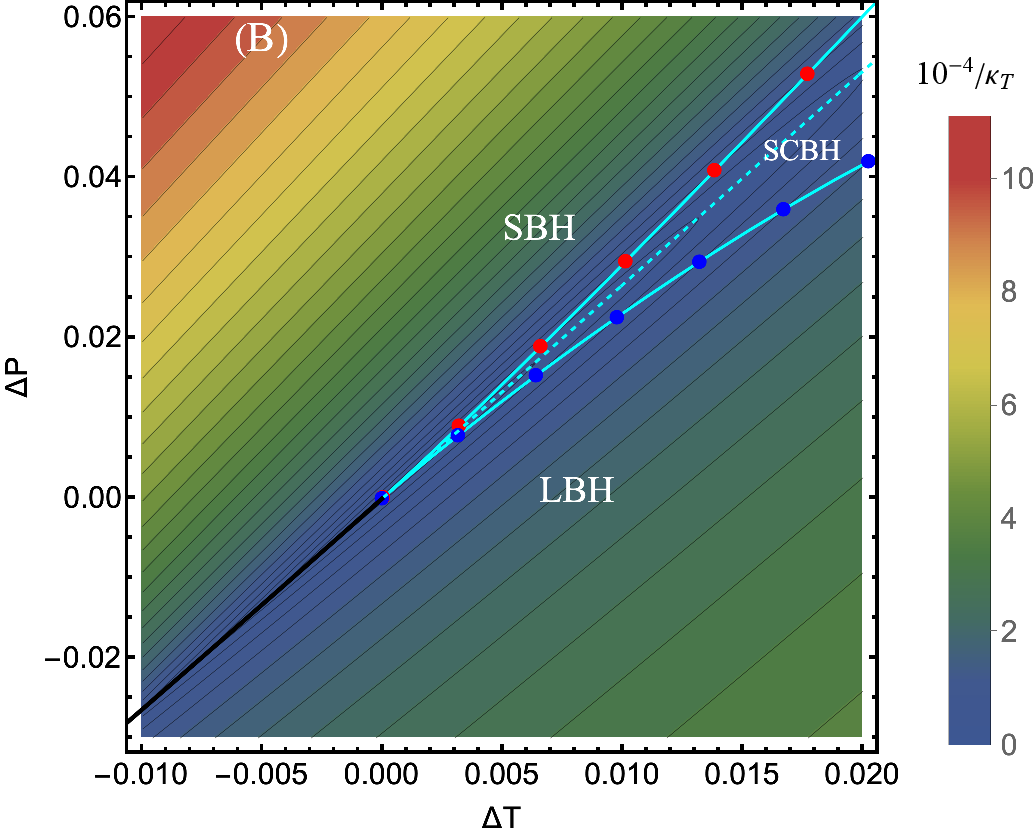}
    \caption{Phase diagrams corresponding to the noncommutative charged AdS black hole in the (A) non-extended phase space, (B) extended phase space. The solid black and dashed cyan lines represent respectively the coexistence line and critical isochore. The solid black point marks the critical point. The solid cyan line with red (blue) points represents the $L^+$ ($L^-$) line. The color map and contour lines are obtained according to the corresponding thermodynamic response function indicated above the color bar. SBH: small black hole, LBH: large black hole, SCBH: SBH-LBH-indistinguishable (or supercritical BH). $\alpha=0.001$.
 }
    \label{fig:contour} 
\end{figure}

To further verify the scaling behavior obtained from the Landau series expansion, we perform numerical calculations using the supercritical crossover lines $L^\pm$. In the non-extended phase space, $T$ is the ordering field, $Q$ is the control field, and $S$ is the order parameter. One first defines the critical isochore as an extension of the coexistence line to the supercritical region: $\delta T(S, Q) = T(S,Q) - T(S_c,Q)$, then the specific heat $C_Q=T\frac{\partial S}{\partial T}|_Q$ is considered as a function of $\Delta Q$ for a few fixed $\delta T(S, Q)$. $C_Q$ peaks at $Q_{max}^+$ for $\delta T>0$ and peaks at $Q_{max}^-$ for $\delta T<0$ for a fixed $\delta T$. Our numerical calculations confirm that near all critical points, the scaling of the $L^\pm$ lines follows Eqs.~(\ref{eq:H_scaling}) and~(\ref{eq:eta_scaling}). This is illustrated in panel (A) and (C) of Fig.~\ref{fig:scaling}, which includes six supercritical scaling lines derived from three different values of noncommutative parameter $\alpha$, namely $\alpha=0, 0.001, 0.01$. The results clearly show that, despite the variation in $\alpha$, the extracted scaling laws remain essentially identical across all three cases, confirming that the supercritical crossover is governed by the same universal mean‑field exponents regardless of the strength of noncommutative corrections. Furthermore, $L^\pm$ segment the black hole's phase diagram into three characteristic zones: the SBH phase, a SCBH phase where SBH and LBH states are indistinguishable, and the LBH phase. In physical units, increasing the value of $\alpha$ within a limited range shifts the critical point toward the origin and compresses the entire phase diagram into a smaller region in the $P-T$ plane. The reduced phase diagram is shown in Fig.~\ref{fig:contour}\,A, which appears unchanged under variation of $\alpha$ due to our normalization by the critical parameters. Since the reduced phase diagram itself shows no discernible $\alpha-$dependence, we only show the representative case $\alpha=0.001$.

\begin{table}[!htbp]
\centering
\caption{Values of the coefficients $a_\pm$ to $d_\pm$. }
\begin{tabular}{lrrrrrrrr}
\hline
\hline
\qquad \qquad & $a_+$ & $a_-$ & $b_+$ & $b_- $& $c_+$ & $c_-$ & $d_+$ & $d_- $\\
\hline
$\alpha=0$ & 0.107  & 0.105 & 0.009 & 0.010 & 2.260  & 2.185 & 0.164 & 0.168\\
$\alpha=0.001$ & 0.107 & 0.105 & 0.009 & 0.010 & 2.260  & 2.185 & 0.165 & 0.168\\
$\alpha=0.01$ & 0.107 & 0.105 & 0.008 & 0.007 & 2.259  & 2.184 & 0.157 & 0.152\\
\hline
\hline
\end{tabular}
\label{tab:coefficients}
\end{table}


\subsection{Extended Ensemble Supercritical Crossover}

Near the critical point, the macroscopic thermodynamics of the black hole simplifies dramatically and admits an effective description in terms of a Landau-type theory. This approach isolates the dominant fluctuations responsible for critical behaviour and allows us to extract universal information without relying on microscopic details. Within the extended phase space, the appropriate order parameter is chosen to be the deviation of the number density from its critical value,
\[
m = \rho - \rho_c \ ,
\]
which continuously interpolates between the two phases that coexist below the critical temperature. Although the first-order transition disappears above \(T_c\), fluctuations in \(m\) continue to characterize the system and control its response to external perturbations.

The thermodynamic quantity conjugate to this order parameter is the pressure. Accordingly, we define the pressure deviation $\delta P=P(\rho,T)-P(\rho_c,T)$ as the control field that drives density fluctuations, in direct analogy with the magnetic field in ferromagnetic systems. Near criticality, the equation of state can be expanded in powers of \(m\), yielding
\[
\delta P = A\, m\,\tau + B\, m^3 + \cdots \ ,
\]
where \(\tau = T - T_c\) measures the distance from the critical temperature. The absence of a quadratic term reflects the inflection-point structure of the equation of state at criticality. The coefficients \(A\) and \(B\) are determined by derivatives of the pressure
evaluated at the critical point,
\[
A = \left.\frac{\partial^2 P}{\partial \rho\, \partial T}\right|_c = 1+\frac{\alpha }{3 \sqrt{6} Q} \, , \qquad
B = \frac{1}{6}\left.\frac{\partial^3 P}{\partial \rho^3}\right|_c = \frac{2 \sqrt{6} Q}{3\pi }-\frac{20}{9 \pi }\alpha \ .
\]
This expansion may be understood as originating from an effective Landau free-energy functional \(F(m, T)\), whose minimization reproduces the equation of state. Within this framework, the linear term encodes the coupling between temperature fluctuations and density deviations, while the cubic term stabilizes the free energy and governs the onset of nonlinear effects near the critical point.

A key quantity of interest is the isothermal susceptibility, which quantifies how strongly the density responds to infinitesimal changes in pressure at fixed temperature. It plays the role of a generalized compressibility and is defined as
\[
\chi_T = \left( \frac{\partial m}{\partial \delta P} \right)_T
= \frac{1}{A\tau + 3B m^2} \, .
\]

At the critical point, where both \(\tau\) and \(m\) vanish, the susceptibility diverges, signaling the emergence of long-range correlations. In the supercritical regime (\(\tau>0\)), this divergence is replaced by pronounced but finite maxima. The locus of these maxima defines the supercritical crossover lines, which extend the coexistence curve beyond the critical point.

To locate these lines analytically, we extremize \(\chi_T\) with respect to the order parameter \(m\). This procedure yields two symmetrical solutions
\begin{eqnarray}\label{m_pmP_pm}
    m_\pm &=& \pm \left(\frac{\sqrt{\pi } \sqrt{\frac{1}{Q}}}{2^{3/4} \sqrt[4]{3}}+\frac{11 \sqrt{\pi } \alpha  \left(\frac{1}{Q}\right)^{3/2}}{12 \sqrt[4]{2} 3^{3/4}}\right)\tau \nonumber  \\ 
    \delta P_\pm &=& \pm \left(\frac{2 \sqrt[4]{\frac{2}{3}} \sqrt{\pi }}{3 \sqrt{Q}}+\frac{13 \sqrt{\pi } \alpha }{9 \sqrt[4]{2} 3^{3/4} Q^{3/2}} \right)\tau^{\frac{3}{2}} \ .
\end{eqnarray}
The two branches correspond to fluctuations toward densities larger or smaller than \(\rho_c\). Although no phase separation occurs above \(T_c\), these solutions capture the dominant directions along which thermodynamic response functions are maximized.  To see the universal content of these results, it is useful to introduce reduced variables
\begin{eqnarray}
    \widetilde\Delta P = \frac{\delta P}{P_c}\ ,
\end{eqnarray}
noting that $P_c \neq P(\rho_c,T)$.
By putting this in Eq.~\eqref{m_pmP_pm}, we have 
\begin{eqnarray}\label{Extended_fan_structure}
\widetilde\Delta P_\pm(\Delta T) = \pm \left(\frac{32\sqrt{6}}{27}  +\frac{172}{81 Q} \alpha \right)\, \Delta T^{3/2} + \cdots \ ,
\end{eqnarray}
This expression reflects the mean-field universality of the
supercritical crossover. Finally, combining Eq.~\eqref{eq:universal_widom_line} and Eq.~\eqref{Extended_fan_structure}, the full structure of the extended-ensemble crossover lines is
\begin{eqnarray}
    \Delta P_{L^\pm}(\Delta T) = \left(\frac{8}{3}+\frac{5 \sqrt{6}}{27 Q} \alpha \right)\,\Delta T \pm \left( \frac{32\sqrt{6}}{27}  +\frac{172}{81 Q} \alpha \right)\,\Delta T^{3/2} + \cdots \ .
\end{eqnarray}

\begin{figure}[th]
    \centering
    \includegraphics[scale=0.55]{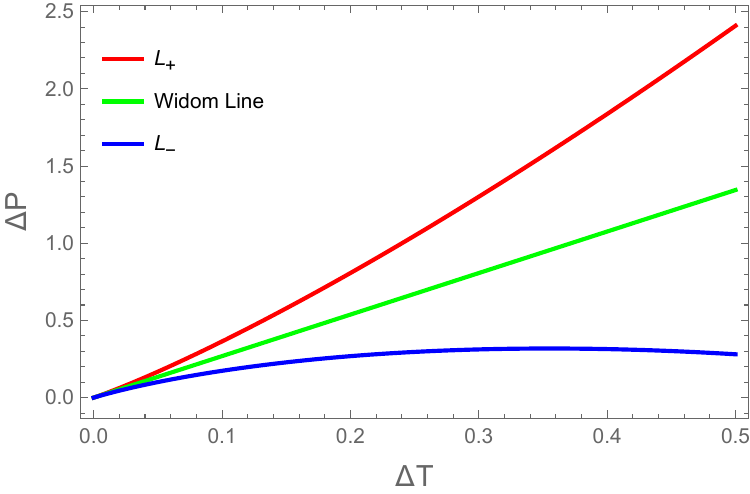}
    \includegraphics[scale=0.55]{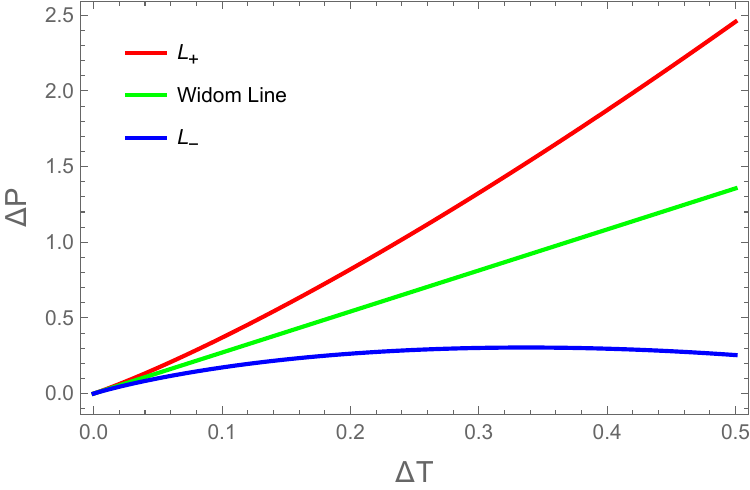}
    \caption{Supercritical crossover structure (\textbf{left : }$\alpha=0.05$ and \textbf{right : }$\alpha=0.1$) in the extended ensemble. This illustrates the crossover fan emerging from the critical point, with its central line corresponding to the Widom line.}
    \label{fig:Extended} 
\end{figure}

The linear contribution represents the smooth thermodynamic background, while the nonanalytic correction captures the leading effect of critical fluctuations. The two branches symmetrically enclose a wedge-shaped region—the crossover fan—within which response functions such as compressibility and specific heat develop pronounced maxima. The central trajectory passing through this fan corresponds to the Widom line, providing a continuous extension of the first-order coexistence curve into the supercritical domain. Figure~\ref{fig:Extended} illustrates the supercritical crossover structure in the extended ensemble for $\alpha=0.05$ and $\alpha=0.1$. The central green curve represents the Widom line, which follows the linear relation $\Delta P = (\frac{8}{3}+\frac{5\sqrt{6}}{27Q}\alpha)\Delta T$ at leading order. The red and blue branches are the two crossover lines obtained from the extremal condition of the isothermal compressibility, which scale as $\Delta P\sim |\Delta T|^{3/2}$ near the critical point. The region enclosed by the two branches forms a wedge‑shaped crossover fan, within which thermodynamic response functions (compressibility) exhibit pronounced maxima. Comparing the left and right panels, one observes that noncommutative effects preserve the universality class while introducing quantitative corrections to the critical amplitudes. As in non-extended ensemble, we also numerically verified the scaling laws of $\delta P\sim |\Delta T|^{3/2}$ and $\delta \rho\sim |\Delta T|^{1/2}$ using supercritical crossover lines, which is illustrated in panel (B) and (D) of Fig.~\ref{fig:scaling}. The data points, obtained for three different values of the noncommutative parameter $\alpha=(0, 0.001, 0.01)$, follow the universal power law, which confirms that the universal mean‑field scaling remains intact regardless of the strength of noncommutative corrections. The corresponding phase diagram in the extended ensemble is shown in Fig.~\ref{fig:contour}\,B, where the coexistence line, critical isochore, and the $L^\pm$ branches are displayed together with a color map of the isothermal compressibility. This figure further demonstrates that the universal supercritical crossover structure, including the crossover fan and the Widom line, persists under noncommutative deformations, and that the reduced phase diagram remains unchanged upon variation of $\alpha$ after normalization by the critical parameters.


\section{Summary and Discussions}\label{sec_summary}
In this work, we have presented a systematic analytical study of the Widom line and the supercritical crossover structure of noncommutative charged AdS black holes. Treating the noncommutative parameter $\alpha$ perturbatively, we derived the corrected thermodynamic quantities, critical points, and the normalized variance $\Omega$ in both the non‑extended (canonical) and extended ensembles. Using the extremal condition of $\Omega$, we obtained closed‑form expressions for the Widom line in each ensemble, showing explicitly how noncommutative effects shift the critical data and the absolute location of the Widom line.

By performing a Landau‑type expansion of the thermodynamic potential near the critical point, we further derived the two symmetric crossover branches $L^\pm$ that emerge from the critical point and extend into the supercritical regime. In the non‑extended ensemble, these branches satisfy $\delta T\sim \left|\Delta Q\right|^{3/2}$, while in the extended ensemble they follow $\delta P\sim \left|\Delta T\right|^{3/2}$. These scaling relations are precisely the mean‑field predictions, and they remain unchanged by the noncommutative deformation. Our numerical verification, using the supercritical crossover lines  $L^{\pm}$ for three representative values of 
$\alpha$, confirms the universal scaling laws and the robustness of the mean‑field description. The supercritical phase diagrams clearly delineate the coexistence line, the critical isochore, and the $L^{\pm}$ branches, together with color maps of the corresponding response functions. These diagrams segment the supercritical black holes into the SBH phase, the LBH phase, and indistinguishable SCBH phase.

Several promising directions remain for future investigation. It would be interesting to explore the influence of higher‑order noncommutative corrections on the supercritical scaling, as well as to extend the analysis to rotating noncommutative black holes or to higher‑curvature theories such as Gauss‑Bonnet gravity. In addition, the dynamical signatures of the Widom line, through quasi‑normal modes or Lee-Yang zeros, could provide a deeper understanding of critical slowing down in noncommutative backgrounds. Finally, interpreting the noncommutative deformation in the dual field theory via the AdS/CFT correspondence may offer a novel perspective on how quantum spacetime effects manifest in strongly coupled systems.


\section*{Acknowledgements}
Ankit Anand is financially supported by the Institute's postdoctoral fellowship at IIT Kanpur. 


\appendix



\bibliographystyle{utphys.bst}
\bibliography{ref}

@article{Anand:2025ab,
    author = "Anand, Ankit and Singh, Aditya and Mishra, Anshul and Channuie, Phongpichit",
    title = "{From Thermodynamics to Dynamics: Perturbative Study in Noncommutative Black Holes}",
    journal = "Submitted to PRD",
    year = "2025"
}

@article{Wang:2025ctk,
    author = "Wang, Shoucheng and Li, Xinyang and Jin, Yuliang and Li, Li",
    title = "{Analogous supercritical crossovers in black holes and water}",
    eprint = "2506.10808",
    archivePrefix = "arXiv",
    primaryClass = "gr-qc",
    month = "6",
    year = "2025"
}

@article{Zhao:2025ecg,
    author = "Zhao, Zi-Qiang and Nie, Zhang-Yu and Zhang, Jing-Fei and Zhang, Xin",
    title = "{Characterized Behaviors of Black Hole Thermodynamics in the Supercritical Region}",
    eprint = "2504.04995",
    archivePrefix = "arXiv",
    primaryClass = "gr-qc",
    doi = "10.1088/0256-307X/42/10/101102",
    journal = "Chin. Phys. Lett.",
    volume = "42",
    number = "10",
    pages = "101102",
    year = "2025"
}

@article{Ouyang:2024ckt,
    author = "Ouyang, Xiao-Yu and Ye, Qi-Jun and Li, Xin-Zheng",
    title = "{Complex phase diagram and supercritical matter}",
    doi = "10.1103/PhysRevE.109.024118",
    journal = "Phys. Rev. E",
    volume = "109",
    number = "2",
    pages = "024118",
    year = "2024"
}

@article{Xu:2025jrk,
    author = "Xu, Zhen-Ming and Mann, Robert B.",
    title = "{Thermodynamic supercriticality and complex phase diagram for the AdS black hole}",
    eprint = "2504.05708",
    archivePrefix = "arXiv",
    primaryClass = "gr-qc",
    month = "4",
    year = "2025"
}

@article{Bokulic:2025mtd,
    author = "Bokuli{\'c}, Ana and Po{\v{z}}ar, Filip",
    title = "{Noncommutative dyonic black holes sourced by nonlinear electromagnetic fields}",
    eprint = "2511.01020",
    archivePrefix = "arXiv",
    primaryClass = "gr-qc",
    reportNumber = "RBI-ThPhys-2025-44",
    doi = "10.1103/t8k4-1thf",
    journal = "Phys. Rev. D",
    volume = "113",
    number = "8",
    pages = "084026",
    year = "2026"
}

@article{Hadri:2025mvu,
    author = "Hadri, Wijdane El and Jemri, Maryem",
    title = "{Thermodynamics and Criticality of Noncommutative RN-AdS Black Holes}",
    eprint = "2509.00926",
    archivePrefix = "arXiv",
    primaryClass = "hep-th",
    month = "8",
    year = "2025"
}

@article{Anand:2025rzh,
    author = "Anand, Ankit and Wang, Shoucheng",
    title = "{Universal Supercritical Behavior in Global Monopole-Charged AdS Black Holes}",
    eprint = "2512.12723",
    archivePrefix = "arXiv",
    primaryClass = "hep-th",
    month = "12",
    year = "2025"
}

@article{Seiberg:1999vs,
    author = "Seiberg, Nathan and Witten, Edward",
    title = "{String theory and noncommutative geometry}",
    eprint = "hep-th/9908142",
    archivePrefix = "arXiv",
    reportNumber = "IASSNS-HEP-99-74",
    doi = "10.1088/1126-6708/1999/09/032",
    journal = "JHEP",
    volume = "09",
    pages = "032",
    year = "1999"
}

@article{Xiao:2023lap,
    author = "Xiao, Yong and Tian, Yu and Liu, Yu-Xiao",
    title = "{Extended Black Hole Thermodynamics from Extended Iyer-Wald Formalism}",
    eprint = "2308.12630",
    archivePrefix = "arXiv",
    primaryClass = "gr-qc",
    doi = "10.1103/PhysRevLett.132.021401",
    journal = "Phys. Rev. Lett.",
    volume = "132",
    number = "2",
    pages = "021401",
    year = "2024"
}

@article{Dolan:2010ha,
    author = "Dolan, Brian P.",
    title = "{The cosmological constant and the black hole equation of state}",
    eprint = "1008.5023",
    archivePrefix = "arXiv",
    primaryClass = "gr-qc",
    reportNumber = "DIAS-STP-10-10",
    doi = "10.1088/0264-9381/28/12/125020",
    journal = "Class. Quant. Grav.",
    volume = "28",
    pages = "125020",
    year = "2011"
}

@article{Kubiznak:2012wp,
    author = "Kubiznak, David and Mann, Robert B.",
    title = "{P-V criticality of charged AdS black holes}",
    eprint = "1205.0559",
    archivePrefix = "arXiv",
    primaryClass = "hep-th",
    doi = "10.1007/JHEP07(2012)033",
    journal = "JHEP",
    volume = "07",
    pages = "033",
    year = "2012"
}

@article{Hawking:1982dh,
    author = "Hawking, S. W. and Page, Don N.",
    title = "{Thermodynamics of Black Holes in anti-De Sitter Space}",
    reportNumber = "PRINT-83-0019 (CAMBRIDGE)",
    doi = "10.1007/BF01208266",
    journal = "Commun. Math. Phys.",
    volume = "87",
    pages = "577",
    year = "1983"
}

@article{10.1063/PT.3.1796,
    author = {Brazhkin, Vadim V. and Trachenko, Kostya},
    title = "{What separates a liquid from a gas?}",
    journal = {Physics Today},
    volume = {65},
    number = {11},
    pages = {68-69},
    year = {2012},
    month = {11},
    issn = {0031-9228},
    doi = {10.1063/PT.3.1796},
}

@article{Xu_2005,
   title={Relation between the Widom line and the dynamic crossover in systems with a liquid–liquid phase transition},
   volume={102},
   ISSN={1091-6490},
   url={http://dx.doi.org/10.1073/pnas.0507870102},
   DOI={10.1073/pnas.0507870102},
   number={46},
   journal={Proceedings of the National Academy of Sciences},
   publisher={Proceedings of the National Academy of Sciences},
   author={Xu, Limei and Kumar, Pradeep and Buldyrev, S. V. and Chen, S.-H. and Poole, P. H. and Sciortino, F. and Stanley, H. E.},
   year={2005},
   month=nov, pages={16558–16562} }

@article{Ruppeiner_2012,
   title={Thermodynamic geometry, phase transitions, and the Widom line},
   volume={86},
   ISSN={1550-2376},
   url={http://dx.doi.org/10.1103/PhysRevE.86.052103},
   DOI={10.1103/physreve.86.052103},
   number={5},
   journal={Physical Review E},
   publisher={American Physical Society (APS)},
   author={Ruppeiner, G. and Sahay, A. and Sarkar, T. and Sengupta, G.},
   year={2012},
   month=nov }

@article{PhysRevLett.112.135701,
  title = {Behavior of the Widom Line in Critical Phenomena},
  author = {Luo, Jiayuan and Xu, Limei and Lascaris, Erik and Stanley, H. Eugene and Buldyrev, Sergey V.},
  journal = {Phys. Rev. Lett.},
  volume = {112},
  issue = {13},
  pages = {135701},
  numpages = {5},
  year = {2014},
  month = {Apr},
  publisher = {American Physical Society},
  doi = {10.1103/PhysRevLett.112.135701},
  url = {https://link.aps.org/doi/10.1103/PhysRevLett.112.135701}
}

@article{PhysRevE.95.052120,
  title = {Similarity law for Widom lines and coexistence lines},
  author = {Banuti, D. T. and Raju, M. and Ihme, M.},
  journal = {Phys. Rev. E},
  volume = {95},
  issue = {5},
  pages = {052120},
  numpages = {5},
  year = {2017},
  month = {May},
  publisher = {American Physical Society},
  doi = {10.1103/PhysRevE.95.052120},
  url = {https://link.aps.org/doi/10.1103/PhysRevE.95.052120}
}

@article{Gallo2014,
  author = {Gallo, P. and Corradini, D. and Rovere, M.},
  title = {Widom line and dynamical crossovers as routes to understand supercritical water},
  journal = {Nature Communications},
  year = {2014},
  volume = {5},
  pages = {5806},
  number = {1},
  issn = {2041-1723},
  url = {https://doi.org/10.1038/ncomms6806},
  doi = {10.1038/ncomms6806},
}

@article{Li_2024,
   title={Thermodynamic crossovers in supercritical fluids},
   volume={121},
   ISSN={1091-6490},
   url={http://dx.doi.org/10.1073/pnas.2400313121},
   DOI={10.1073/pnas.2400313121},
   number={18},
   journal={Proceedings of the National Academy of Sciences},
   publisher={Proceedings of the National Academy of Sciences},
   author={Li, Xinyang and Jin, Yuliang},
   year={2024},
   month=apr }

@article{10.1063/1.1671624,
    author = {Fisher, Michael E. and Wiodm, B.},
    title = "{Decay of Correlations in Linear Systems}",
    journal = {The Journal of Chemical Physics},
    volume = {50},
    number = {9},
    pages = {3756-3772},
    year = {1969},
    month = {05},
    issn = {0021-9606},
    doi = {10.1063/1.1671624},
}

@article{RJFLeotedeCarvalho_1994,
doi = {10.1088/0953-8984/6/44/008},
url = {https://dx.doi.org/10.1088/0953-8984/6/44/008},
year = {1994},
month = {oct},
publisher = {},
volume = {6},
number = {44},
pages = {9275},
author = {R J F Leote de Carvalho and  R Evans and  D C Hoyle and  J R Henderson},
title = {The decay of the pair correlation function in simple fluids: long- versus short-ranged potentials},
journal = {Journal of Physics: Condensed Matter},
}

@article{PhysRevE.51.3146,
  title = {Location of the Fisher-Widom line for systems interacting through short-ranged potentials},
  author = {Vega, C. and Rull, L. F. and Lago, S.},
  journal = {Phys. Rev. E},
  volume = {51},
  issue = {4},
  pages = {3146--3155},
  numpages = {0},
  year = {1995},
  month = {Apr},
  publisher = {American Physical Society},
  doi = {10.1103/PhysRevE.51.3146},
  url = {https://link.aps.org/doi/10.1103/PhysRevE.51.3146}
}

@article{Bolmatov2013,
  author = {Bolmatov, Dima and Brazhkin, V. V. and Trachenko, K.},
  title = {Thermodynamic behaviour of supercritical matter},
  journal = {Nature Communications},
  year = {2013},
  volume = {4},
  pages = {2331},
  number = {1},
  issn = {2041-1723},
  url = {https://doi.org/10.1038/ncomms3331},
  doi = {10.1038/ncomms3331},
}

@article{Yoon_2018,
   title={“Two-Phase” Thermodynamics of the Frenkel Line},
   volume={9},
   ISSN={1948-7185},
   url={http://dx.doi.org/10.1021/acs.jpclett.8b01955},
   DOI={10.1021/acs.jpclett.8b01955},
   number={16},
   journal={The Journal of Physical Chemistry Letters},
   publisher={American Chemical Society (ACS)},
   author={Yoon, Tae Jun and Ha, Min Young and Lee, Won Bo and Lee, Youn-Woo},
   year={2018},
   month=jul, pages={4550–4554} }

@article{PhysRevLett.111.145901,
  title = {``Liquid-Gas'' Transition in the Supercritical Region: Fundamental Changes in the Particle Dynamics},
  author = {Brazhkin, V. V. and Fomin, Yu. D. and Lyapin, A. G. and Ryzhov, V. N. and Tsiok, E. N. and Trachenko, Kostya},
  journal = {Phys. Rev. Lett.},
  volume = {111},
  issue = {14},
  pages = {145901},
  numpages = {5},
  year = {2013},
  month = {Oct},
  publisher = {American Physical Society},
  doi = {10.1103/PhysRevLett.111.145901},
  url = {https://link.aps.org/doi/10.1103/PhysRevLett.111.145901}
}

@article{Prescher_2017,
   title={Experimental evidence of the Frenkel line in supercritical neon},
   volume={95},
   ISSN={2469-9969},
   url={http://dx.doi.org/10.1103/PhysRevB.95.134114},
   DOI={10.1103/physrevb.95.134114},
   number={13},
   journal={Physical Review B},
   publisher={American Physical Society (APS)},
   author={Prescher, C. and Fomin, Yu. D. and Prakapenka, V. B. and Stefanski, J. and Trachenko, K. and Brazhkin, V. V.},
   year={2017},
   month=apr }

@article{Bolmatov_2015,
   title={The Frenkel Line: a direct experimental evidence for the new thermodynamic boundary},
   volume={5},
   ISSN={2045-2322},
   url={http://dx.doi.org/10.1038/srep15850},
   DOI={10.1038/srep15850},
   number={1},
   journal={Scientific Reports},
   publisher={Springer Science and Business Media LLC},
   author={Bolmatov, Dima and Zhernenkov, Mikhail and Zav’yalov, Dmitry and Tkachev, Sergey N. and Cunsolo, Alessandro and Cai, Yong Q.},
   year={2015},
   month=nov }

@article{Fomin_2018,
doi = {10.1088/1361-648X/aaaf39},
url = {https://dx.doi.org/10.1088/1361-648X/aaaf39},
year = {2018},
month = {mar},
publisher = {IOP Publishing},
volume = {30},
number = {13},
pages = {134003},
author = {Yu D Fomin and V N Ryzhov and E N Tsiok and J E Proctor and C Prescher and V B Prakapenka and K Trachenko and V V Brazhkin},
title = {Dynamics, thermodynamics and structure of liquids and supercritical fluids: crossover at the Frenkel line},
journal = {Journal of Physics: Condensed Matter},
}

@article{Fomin2015,
  author = {Fomin, Yu. D. and Ryzhov, V. N. and Tsiok, E. N. and Brazhkin, V. V.},
  title = {Dynamical crossover line in supercritical water},
  journal = {Scientific Reports},
  year = {2015},
  volume = {5},
  pages = {14234},
  number = {1},
  issn = {2045-2322},
  url = {https://doi.org/10.1038/srep14234},
  doi = {10.1038/srep14234},
}

@article{PhysRevE.85.031203,
  title = {Two liquid states of matter: A dynamic line on a phase diagram},
  author = {Brazhkin, V. V. and Fomin, Yu. D. and Lyapin, A. G. and Ryzhov, V. N. and Trachenko, K.},
  journal = {Phys. Rev. E},
  volume = {85},
  issue = {3},
  pages = {031203},
  numpages = {12},
  year = {2012},
  month = {Mar},
  publisher = {American Physical Society},
  doi = {10.1103/PhysRevE.85.031203},
  url = {https://link.aps.org/doi/10.1103/PhysRevE.85.031203}
}

@article{2023PhRvR...5a3149H,
       author = {{Huang}, Dong and {Baggioli}, Matteo and {Lu}, Shaoyu and {Ma}, Zhuang and {Feng}, Yan},
        title = "{Revealing the supercritical dynamics of dusty plasmas and their liquidlike to gaslike dynamical crossover}",
      journal = {Physical Review Research},
         year = 2023,
        month = feb,
       volume = {5},
       number = {1},
          eid = {013149},
        pages = {013149},
          doi = {10.1103/PhysRevResearch.5.013149},
archivePrefix = {arXiv},
       eprint = {2301.08449},
 primaryClass = {physics.plasm-ph},
       adsurl = {https://ui.adsabs.harvard.edu/abs/2023PhRvR...5a3149H}
}

@article{PhysRevLett.75.1040,
  title = {Probing the Nuclear Liquid-Gas Phase Transition},
  author = {Pochodzalla, J. and M\"ohlenkamp, T. and Rubehn, T. and Sch\"uttauf, A. and W\"orner, A. and Zude, E. and Begemann-Blaich, M. and Blaich, Th. and Emling, H. and Ferrero, A. and Gross, C. and Imm\'e, G. and Iori, I. and Kunde, G. J. and Kunze, W. D. and Lindenstruth, V. and Lynen, U. and Moroni, A. and M\"uller, W. F. J. and Ocker, B. and Raciti, G. and Sann, H. and Schwarz, C. and Seidel, W. and Serfling, V. and Stroth, J. and Trautmann, W. and Trzcinski, A. and Tucholski, A. and Verde, G. and Zwieglinski, B.},
  journal = {Phys. Rev. Lett.},
  volume = {75},
  issue = {6},
  pages = {1040--1043},
  numpages = {0},
  year = {1995},
  month = {Aug},
  publisher = {American Physical Society},
  doi = {10.1103/PhysRevLett.75.1040},
  url = {https://link.aps.org/doi/10.1103/PhysRevLett.75.1040}
}

@article{Cvetic_1999,
   title={Phases of R-charged black holes, spinning branes and strongly coupled gauge theories},
   volume={1999},
   ISSN={1029-8479},
   url={http://dx.doi.org/10.1088/1126-6708/1999/04/024},
   DOI={10.1088/1126-6708/1999/04/024},
   number={04},
   journal={Journal of High Energy Physics},
   publisher={Springer Science and Business Media LLC},
   author={Cvetic, Mirjam and Gubser, Steven S},
   year={1999},
   month=apr, pages={024–024} }

@article{Cvetic:1999rb,
    author = "Cvetic, Mirjam and Gubser, Steven S.",
    title = "{Thermodynamic stability and phases of general spinning branes}",
    eprint = "hep-th/9903132",
    archivePrefix = "arXiv",
    reportNumber = "HUTP-A086, UPR-826-T",
    doi = "10.1088/1126-6708/1999/07/010",
    journal = "JHEP",
    volume = "07",
    pages = "010",
    year = "1999"
}

@article{Jim_nez_2021,
   title={A quantum magnetic analogue to the critical point of water},
   volume={592},
   ISSN={1476-4687},
   url={http://dx.doi.org/10.1038/s41586-021-03411-8},
   DOI={10.1038/s41586-021-03411-8},
   number={7854},
   journal={Nature},
   publisher={Springer Science and Business Media LLC},
   author={Jiménez, J. Larrea and Crone, S. P. G. and Fogh, E. and Zayed, M. E. and Lortz, R. and Pomjakushina, E. and Conder, K. and Läuchli, A. M. and Weber, L. and Wessel, S. and Honecker, A. and Normand, B. and Rüegg, Ch. and Corboz, P. and Rønnow, H. M. and Mila, F.},
   year={2021},
   month=apr, pages={370–375} }

@article{doi:10.1021/acs.jpcb.9b04058,
author = {Ploetz, Elizabeth
A. and Smith, Paul E.},
title = {Gas or Liquid? The Supercritical Behavior of Pure Fluids},
journal = {The Journal of Physical Chemistry B},
volume = {123},
number = {30},
pages = {6554-6563},
year = {2019},
doi = {10.1021/acs.jpcb.9b04058},
    note ={PMID: 31287691},
}

@article{jiang2024experimentalobservationgappedshear,
      title={Experimental observation of gapped shear waves and liquid-like to gas-like dynamical crossover in active granular matter}, 
      author={Cunyuan Jiang and Zihan Zheng and Yangrui Chen and Matteo Baggioli and Jie Zhang},
      year={2024},
      eprint={2403.08285},
      archivePrefix={arXiv},
      primaryClass={cond-mat.soft},
      url={https://arxiv.org/abs/2403.08285}, 
}

@article{Stephanov:2008qz,
    author = "Stephanov, M. A.",
    title = "{Non-Gaussian fluctuations near the QCD critical point}",
    eprint = "0809.3450",
    archivePrefix = "arXiv",
    primaryClass = "hep-ph",
    doi = "10.1103/PhysRevLett.102.032301",
    journal = "Phys. Rev. Lett.",
    volume = "102",
    pages = "032301",
    year = "2009"
}

@article{Stephanov:2011pb,
    author = "Stephanov, M. A.",
    title = "{On the sign of kurtosis near the QCD critical point}",
    eprint = "1104.1627",
    archivePrefix = "arXiv",
    primaryClass = "hep-ph",
    doi = "10.1103/PhysRevLett.107.052301",
    journal = "Phys. Rev. Lett.",
    volume = "107",
    pages = "052301",
    year = "2011"
}

@article{Sordi:2023cjq,
    author = "Sordi, G. and Tremblay, A. -M. S.",
    title = "Introducing the concept of the Widom line in the QCD phase diagram",
    eprint = "2312.12401",
    archivePrefix = "arXiv",
    primaryClass = "hep-ph",
    doi = "10.1103/PhysRevD.109.114020",
    journal = "Phys. Rev. D",
    volume = "109",
    number = "11",
    pages = "114020",
    year = "2024"
}

@article{DasBairagya:2019nyv,
    author = "Das Bairagya, Joy and Pal, Kunal and Pal, Kuntal and Sarkar, Tapobrata",
    title = "{The geometry of RN-AdS fluids}",
    eprint = "1912.01183",
    archivePrefix = "arXiv",
    primaryClass = "hep-th",
    doi = "10.1016/j.physletb.2020.135416",
    journal = "Phys. Lett. B",
    volume = "805",
    pages = "135416",
    year = "2020"
}

@article{Sahay:2017hlq,
    author = "Sahay, Anurag and Jha, Rishabh",
    title = "{Geometry of criticality, supercriticality and Hawking-Page transitions in Gauss-Bonnet-AdS black holes}",
    eprint = "1707.03629",
    archivePrefix = "arXiv",
    primaryClass = "hep-th",
    doi = "10.1103/PhysRevD.96.126017",
    journal = "Phys. Rev. D",
    volume = "96",
    number = "12",
    pages = "126017",
    year = "2017"
}

@article{Wei:2019yvs,
    author = "Wei, Shao-Wen and Liu, Yu-Xiao and Mann, Robert B.",
    title = "{Ruppeiner Geometry, Phase Transitions, and the Microstructure of Charged AdS Black Holes}",
    eprint = "1909.03887",
    archivePrefix = "arXiv",
    primaryClass = "gr-qc",
    doi = "10.1103/PhysRevD.100.124033",
    journal = "Phys. Rev. D",
    volume = "100",
    number = "12",
    pages = "124033",
    year = "2019"
}

@article{Wei:2023mxw,
    author = "Wei, Shao-Wen and Liu, Yu-Xiao",
    title = "{Thermodynamic nature of black holes in coexistence region}",
    eprint = "2308.11886",
    archivePrefix = "arXiv",
    primaryClass = "gr-qc",
    doi = "10.1007/s11433-023-2335-2",
    journal = "Sci. China Phys. Mech. Astron.",
    volume = "67",
    number = "5",
    pages = "250412",
    year = "2024"
}

@article{Simeoni2010,
    author = "Simeoni, G. G. and Bryk, T. and Gorelli, F. A. and Krisch, M. and Ruocco, G. and Santoro, M. and Scopigno, T.",
    title = "The Widom line as the crossover between liquid-like and gas-like behaviour in supercritical fluids",
    journal = "Nature Physics",
    volume = "6",
    number = "7",
    pages = "503-507",
    year = "2010",
    doi = "10.1038/nphys1683",
    url = "https://doi.org/10.1038/nphys1683"
}

@book{stanley1971phase,
  title={Introduction to Phase Transitions and Critical Phenomena},
  author={Stanley, H. Eugene},
  year={1971},
  publisher={Oxford University Press},
  address={Oxford}
}

@book{goldenfeld1992lectures,
  title={Lectures on Phase Transitions and the Renormalization Group},
  author={Goldenfeld, Nigel},
  year={1992},
  publisher={Addison-Wesley},
  address={Reading, MA}
}

@article{kubiznak2012pvn,
  title={{P–V} criticality of charged AdS black holes},
  author={Kubizňák, David and Mann, Robert B.},
  journal={Journal of High Energy Physics},
  volume={2012},
  number={7},
  pages={33},
  year={2012},
  publisher={Springer},
  doi={10.1007/JHEP07(2012)033}
}

@article{gunasekaran2012extended,
  title={Extended phase space thermodynamics for charged and rotating black holes and Born–Infeld vacuum polarization},
  author={Gunasekaran, Sharmila and Kubizňák, David and Mann, Robert B.},
  journal={Journal of High Energy Physics},
  volume={2012},
  number={11},
  pages={110},
  year={2012},
  publisher={Springer},
  doi={10.1007/JHEP11(2012)110}
}

@article{wei2019critical,
  title={Critical phenomena and thermodynamic geometry of charged Gauss–Bonnet AdS black holes},
  author={Wei, Shao-Wen and Liu, Yu-Xiao},
  journal={Physical Review D},
  volume={100},
  number={8},
  pages={084057},
  year={2019},
  publisher={APS},
  doi={10.1103/PhysRevD.100.084057}
}

@article{Simeski2023,
    author = "Simeski, Filip and Ihme, Matthias",
    title = "{Supercritical fluids behave as complex networks}",
    journal = "Nature Communications",
    volume = "14",
    number = "1",
    pages = "1996",
    year = "2023",
    doi = "10.1038/s41467-023-37645-z",
    url = "https://doi.org/10.1038/s41467-023-37645-z"
}

@article{wei2020critical,
  title={Critical phenomena in charged AdS black holes: reentrant phase transitions and triple points},
  author={Wei, Xian-Neng and Liu, Yong},
  journal={Phys. Rev. D},
  volume={101},
  pages={024009},
  year={2020},
  doi={10.1103/PhysRevD.101.024009}
}

@article{altamirano2013reentrant,
  title={Reentrant phase transitions in rotating anti–de Sitter black holes},
  author={Altamirano, N. and Kubiznak, D. and Mann, R. B.},
  journal={Phys. Rev. D},
  volume={88},
  pages={101502},
  year={2013},
  doi={10.1103/PhysRevD.88.101502}
}

@article{abbott2016gw150914,
  title={Observation of Gravitational Waves from a Binary Black Hole Merger},
  author={Abbott, B. P. and et al.},
  journal={Phys. Rev. Lett.},
  volume={116},
  pages={061102},
  year={2016},
  doi={10.1103/PhysRevLett.116.061102},
  eprint={1602.03837}
}

@article{abbott2021gwtc2,
  title={GWTC-2: Compact Binary Coalescences Observed by LIGO and Virgo During the First Half of the Third Observing Run},
  author={Abbott, B. P. and et al.},
  journal={Phys. Rev. X},
  volume={11},
  pages={021053},
  year={2021},
  doi={10.1103/PhysRevX.11.021053},
  eprint={2010.14527}
}

@article{domenech2021cosmogw,
  title={Cosmological gravitational waves from isocurvature fluctuations},
  author={Domenech, G. and others},
  journal={Universe},
  volume={7},
  pages={398},
  year={2021},
  doi={10.3390/universe7120398}
}

@article{nicola2010nc,
  title={Noncommutative geometry inspired Schwarzschild black holes},
  author={Nicolini, P.},
  journal={Int. J. Mod. Phys. A},
  volume={24},
  pages={1229--1308},
  year={2010},
  doi={10.1142/S0217751X10049421}
}

@article{smeared2006nc,
  title={Smeared quantum black holes in noncommutative geometry},
  author={Smailagic, A. and Spallucci, E.},
  journal={J. Phys. A: Math. Gen.},
  volume={36},
  pages={L467--L471},
  year={2003},
  doi={10.1088/0305-4470/36/33/L02}
}

@article{smailagic2003fuzzy,
  title={Feynman path integral on the non-commutative plane},
  author={Smailagic, A. and Spallucci, E.},
  journal={J. Phys. A: Math. Gen.},
  volume={36},
  pages={L517--L521},
  year={2003},
  doi={10.1088/0305-4470/36/33/L05}
}

@article{nozari2008nc,
  title={Some aspects of Planck scale quantum optics},
  author={Nozari, K. and Mehdipour, S. H.},
  journal={Chaos, Solitons and Fractals},
  volume={34},
  pages={224--231},
  year={2007},
  doi={10.1016/j.chaos.2006.06.045}
}

@article{gonin2025pbh,
  title={Primordial Black-Hole Formation Heavy r-process Element Synthesis from the QCD Transition},
  author={Gonin, M. and Ropke, G.},
  journal={Eur. Phys. J. A},
  volume={61},
  pages={91},
  year={2025},
  doi={10.1140/epja/s10050-025-01639-w}
}

@article{hartnoll2008holographic,
  title={Building a holographic superconductor},
  author={Hartnoll, S. A. and Herzog, C. P. and Horowitz, G. T.},
  journal={Phys. Rev. Lett.},
  volume={101},
  pages={031601},
  year={2008},
  doi={10.1103/PhysRevLett.101.031601}
}

@article{maldacena1998largeN,
  title={The large-N limit of superconformal field theories and supergravity},
  author={Maldacena, J. M.},
  journal={Adv. Theor. Math. Phys.},
  volume={2},
  pages={231--252},
  year={1998},
  eprint={hep-th/9711200}
}

@article{witten1998ads,
  title={Anti de Sitter space and holography},
  author={Witten, E.},
  journal={Adv. Theor. Math. Phys.},
  volume={2},
  pages={253--291},
  year={1998},
  eprint={hep-th/9802150}
}

@article{aharony2000largeN,
  title={Large N field theories, string theory and gravity},
  author={Aharony, O. and Gubser, S. S. and Maldacena, J. and Ooguri, H. and Oz, Y.},
  journal={Phys. Rep.},
  volume={323},
  pages={183--386},
  year={2000},
  doi={10.1016/S0370-1573(99)00083-6}
}

@article{gubser1998gauge,
  title={Gauge theory correlators from non‑critical string theory},
  author={Gubser, S. S. and Klebanov, I. R. and Polyakov, A. M.},
  journal={Phys. Lett. B},
  volume={428},
  pages={105--114},
  year={1998},
  eprint={hep-th/9802109}
}

@article{witten1998confinement,
  title={Anti de Sitter space, thermal phase transition, and confinement in gauge theories},
  author={Witten, E.},
  journal={Adv. Theor. Math. Phys.},
  volume={2},
  pages={505--532},
  year={1998},
  eprint={hep-th/9803131}
}

@article{15,
  author    = {N. Seiberg and E. Witten},
  title     = {String Theory and Noncommutative Geometry},
  journal   = {JHEP},
  volume    = {9909},
  pages     = {032},
  year      = {1999},
  eprint    = {hep-th/9908142},
  archivePrefix = {arXiv}
}

@article{160,
  author    = {M. K. Parikh and F. Wilczek},
  title     = {Hawking Radiation as Tunneling},
  journal   = {Phys. Rev. Lett.},
  volume    = {85},
  pages     = {5042},
  year      = {2000}
}

@article{Kastor:2009wy,
    author = "Kastor, David and Ray, Sourya and Traschen, Jennie",
    title = "{Enthalpy and the Mechanics of AdS Black Holes}",
    eprint = "0904.2765",
    archivePrefix = "arXiv",
    primaryClass = "hep-th",
    doi = "10.1088/0264-9381/26/19/195011",
    journal = "Class. Quant. Grav.",
    volume = "26",
    pages = "195011",
    year = "2009"
}

@article{Kubiznak:2016qmn,
    author = "Kubiznak, David and Mann, Robert B. and Teo, Mae",
    title = "{Black hole chemistry: thermodynamics with Lambda}",
    eprint = "1608.06147",
    archivePrefix = "arXiv",
    primaryClass = "hep-th",
    doi = "10.1088/1361-6382/aa5c69",
    journal = "Class. Quant. Grav.",
    volume = "34",
    number = "6",
    pages = "063001",
    year = "2017"
}


\appendix

\end{document}